\documentclass[aps, twocolumn, longbibliography]{revtex4-2}

\usepackage[colorlinks,urlcolor=blue,citecolor=blue,linkcolor=blue]{hyperref}

\usepackage{amsmath}
\usepackage{amsfonts}
\usepackage{graphicx}
\usepackage{yhmath}
\usepackage{physics}
\usepackage{bm}
\usepackage{mathdots}
\usepackage{MnSymbol}

\newcommand{\D}{\mathbf{D}}
\newcommand{\W}{\mathbf{W}}

\begin{document}

\title{Solving Distance-Based Optimization Problems Using Optical Hardware}

\author{Guangyao Li}	
\email[correspondence address: ]{gl559@cam.ac.uk}
\affiliation{Department of Applied Mathematics and Theoretical Physics, University of Cambridge, Cambridge CB3 0WA, United Kingdom}

\author{Richard Zhipeng Wang}	
\affiliation{Department of Applied Mathematics and Theoretical Physics, University of Cambridge, Cambridge CB3 0WA, United Kingdom}

\author{Natalia G. Berloff}	
\affiliation{Department of Applied Mathematics and Theoretical Physics, University of Cambridge, Cambridge CB3 0WA, United Kingdom}

%\date{\today}

\begin{abstract}
We present a practical approach to solving distance-based optimization problems using optical computing hardware. The objective is to minimize an energy function defined as the weighted sum of squared differences between measured distances and the squared Euclidean distances of point coordinates. By representing coordinates as complex numbers, we map the optimization problem onto optical fields, enabling its solution through either the canonical transformation (CT) method or the gain-based bifurcation (GBB) method. To further enhance the performance of the CT method, we introduce two techniques: asynchronous update and steepened gradient. Both the CT and GBB methods can effectively solve the distance-based problem and are adaptable to various optical hardware platforms. Our optical implementation is inspired by recent progress in analog optical computing for combinatorial optimization, highlighting its promise for efficient and scalable problem-solving in high-dimensional settings.
\end{abstract}
\maketitle

\section{Introduction}\label{sec:introduction}

The field of computational optimization has witnessed a paradigm shift with the advent of specialized computing architectures designed to tackle optimization challenges and enhance machine learning processes~\cite{Li2021,Nikita_AQT2023}. These architectures exploit various physical principles to guide systems toward equilibrium states efficiently. The underlying principles range from thermal and quantum annealing~\cite{Mohseni_REV2022} to Hopf bifurcation at the condensation threshold~\cite{Strogatz2015,Marvin_2023}, minimum power dissipation~\cite{Parrondo_REV2015,Yablonovitch_PNAS2020}, and the principle of least action~\cite{feynman1964lectures}. These principles form the foundation of efficient physics-inspired computing heuristics that are collectively known as pi- ($\pi$-) computing. 

Coupled light-matter complex-valued neural networks represent a particularly innovative approach to computing, distinct from traditional gate-based methods and annealing techniques, both quantum and classical. Exciton-polariton condensates~\cite{Deng_RMP2010,Byrnes2014} epitomize the coupling of light and matter, giving rise to Gain-Based Computing (GBC)~\cite{Opala_REV2023,cummins2023}. GBC leverages the interplay of light and matter, employing advanced laser technologies and spatial light modulators (SLMs) to facilitate parallel processing across numerous channels. This approach is assisted by the nonlinearities intrinsic to the matter component, achieving exceptional energy efficiency~\cite{Opala_PRApplied2019,Opala_REV2023}.

The operational sequence of GBC involves a unique coupling of increased gain power with symmetry-breaking and gradient-descent mechanisms. This process harnesses the inherent coherence and synchronization of waves, naturally guiding the system towards a state of minimized losses. By fully exploiting light's degrees of freedom (both amplitudes and phases) these systems potentially enable richer computational capabilities in photonic or polaritonic complex-valued neural networks~\cite{Sedov_2025}, enhancing their ability to process and interpret complex data patterns efficiently.

Recent advancements in GBC-inspired hardware have introduced a variety of technologies capable of simulating spin Hamiltonians. These include coherent Ising machines based on optical parametric oscillators (OPOs)~\cite{Peter_Science2016,Yamamoto_REV2020,Gao_REV2024}, advanced laser systems~\cite{Shastri2021,WU2022133}, photonic simulators~\cite{Prabhu_Optica2020}, and both polariton and photon condensates~\cite{Berloff2017,Peng_NanoPho2024}. These approaches primarily focus on solving optimization problems on sets of either discrete or continuous variables, which can be mapped onto the phases of generally complex signal amplitudes. The goal is typically to find the ground state of various classical spin Hamiltonians such as Ising, XY~\cite{Berloff2017}, Clock~\cite{Zhang_PRE2020}, and k-local models~\cite{Bairey_PRL2019}.

While these physical platforms have shown great promise, their current focus on problems that can be mapped into standard spin Hamiltonians restricts the scope of potential applications. Recognizing this limitation, researchers have proposed several extensions to broaden the class of solvable problems. These extensions aim to encompass mixed-integer optimization problems~\cite{kalinin2023}, box-constrained quadratic optimization problems~\cite{khosravi2023, kamaletdinov2024coupling} and phase retrieval problems \cite{Wang2025}, potentially opening up new avenues for solving a wider range of complex optimization challenges using physical systems.

In this paper we extend the range of problems that can be efficiently solved using optical hardware to a wide class of distance-based optimization problems \cite{Dokmani_REV2015}. Our paper is organised as follows. In Section \ref{sec:distance_based} we present the problem formulation and discuss various real-life problems. In Section \ref{sec:proposed_optical_implementation} we discuss the proposed applications using the system of coupled oscillators. The numerical results are presented in Section \ref{sec:numerical_result}. We conclude in Section \ref{sec:conclusions}.

\section{Distance Based Formulation}\label{sec:distance_based}

%\subsection{Optimization Problems}

Optimization problems involving pairwise distances are ubiquitous in science and engineering~\cite{Liberti_DistanceREV}. These problems often involve minimizing an energy function of the form
\begin{equation}
E = \sum_{ij} w_{ij} (\norm{\mathbf{x}_i - \mathbf{x}_j}^2 - d_{ij}^2)^2,
\end{equation}
where $\mathbf{x}_i$ denotes the coordinates of point (or object) $i$, $d_{ij}$ is the target distance between points $i$ and $j$, and $w_{ij} \ge 0$ is a weight reflecting the confidence or importance of the constraint between them. We denote the distance matrix by $\D$ and the weight matrix by $\W$.

Several real-world problems can be formulated using this energy function, while the roles of $w_{ij}$ and $d_{ij}$ in each context could be different. In optical contexts, we deal with the system of oscillators described by complex-valued functions $\psi_i(t)$. The position of each point in two-dimensional space can be mapped onto the real and imaginary parts of $\psi_i=R_i \exp(i \theta_i)$ as $\mathbf{x}_i = (R_i \cos\theta_i, R_i\sin\theta_i)$. This allows us to represent each point in terms of the amplitude $R_i$ and phase $\theta_i$ of each oscillator. The  problem is to minimize
\begin{equation}
E_{2D} = \frac{1}{2}\sum_{i,j} w_{ij} (\abs{\psi_i -\psi_j}^2 - d_{ij}^2)^2, \label{E2}
\end{equation}
where the additional $1/2$ factor is for the convenience of differentiation and does not affect the minimizer values.
Below we list some of the real-life problems that can exploit this approach.

\begin{figure}
  \centering
  \includegraphics[width=0.8\linewidth]{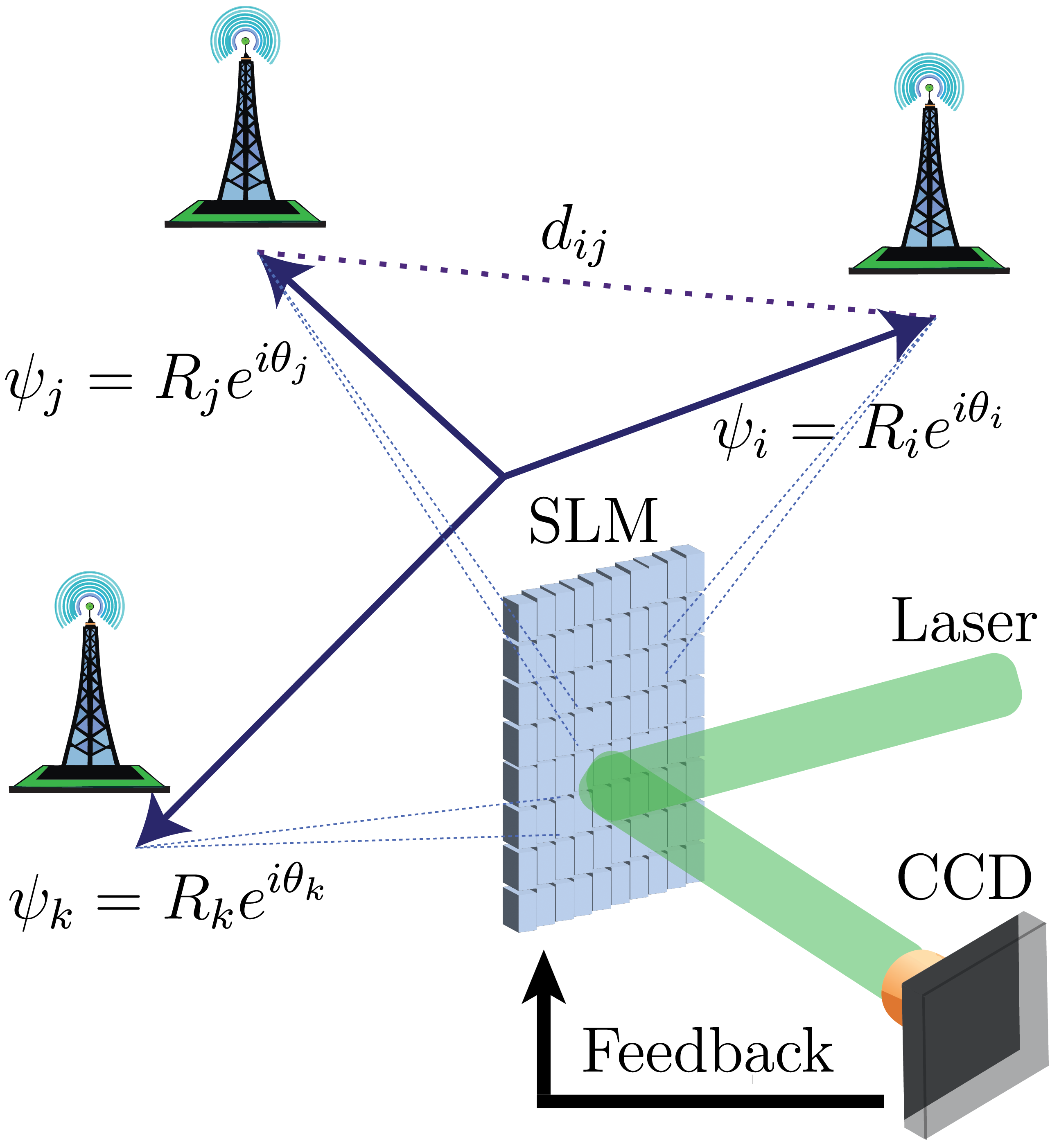}
  \caption{Basic scheme of a spatial photonic machine, where complex numbers are encoded into optical modes using a spatial light modulator (SLM), and interactions are controlled through intensity modulation. Recurrent electronic feedback from the charge-coupled device (CCD) camera guides the configuration toward the ground state.}
  \label{fig:2D3D_illustration}
\end{figure}

\emph{Wireless Sensor Network Localization (SNL).}
In wireless sensor networks~\cite{AspnesIEEE2006}, $\psi_i$ would represent the two-dimensional position $(R_i \cos\theta_i, R_i\sin\theta_i)$ of sensor $i$, $d_{ij}$ is the measured distance between sensors $i$ and $j$ (often based on signal strength), and $w_{ij}$ represents the reliability of the measurement, which may depend on factors such as signal strength and environmental conditions. The standard SNL problem employs a weight matrix $\W$ (also referred to as the mask matrix), where each element $w_{ij} \in \{0,1\}$ signifies whether the measurement is out-of-range or within-range, respectively. Nonetheless, our results easily generalize to the scenario of continuous weights, specifically when $w_{ij} \in [0,1]$.
Figure~\ref{fig:2D3D_illustration} presents a schematic illustration of the optical realization. Practical problems introduce modifications that will make the SNL problem more difficult such as designating some nodes as ``anchors" with a fixed position, or considering noise (inconsistency) in the $\D$ matrix~\cite{Dokmani_REV2015}. Previous studies have shown that the SNL problem is a non-deterministic polynomial-time hard (NP-hard) problem for sparse graphs~\cite{AspnesSNLNP}, making efficient approximate localization algorithms crucial for practical applications.

%%2D3D_illustration

\emph{Social Network Visualization.}
In social network analysis~\cite{Social_network_analysis}, $\psi_i$ would represent the position of individual $i$ in a two-dimensional social space, $d_{ij}$ represents the ``social distance" between individuals (e.g., based on interaction frequency or relationship strength), and $w_{ij}$ might represent the confidence in the measured social distance, possibly based on the amount or quality of data available for that relationship.

\emph{Market Segmentation.}
In the context of market segmentation~\cite{Foedermayr01072008}, $\psi_i$ represents the position of product $i$ in a two-dimensional perceptual map, $d_{ij}$ represents the perceived similarity or difference between products $i$ and $j$ based on consumer data, and $w_{ij}$ might represent the reliability of the similarity measure, possibly based on sample size or consistency of consumer responses.

The distance-based formulation can be extended to objects in higher-dimensional spaces. This is achieved by introducing additional oscillators (represented as complex numbers) to encode the positions of these objects. For instance, in three-dimensional space, we can use $\bm{\psi}_i = (\psi_{1,i}, \operatorname{Re}\psi_{2,i})$, to characterize the position $\mathbf{x}_i=(R_{1,i}\cos\theta_{1,i},R_{1,i}\sin\theta_{1,i}, R_{2,i}\cos\theta_{2,i}),$ where $\psi_{k,i}=R_{k,i} \exp(i \theta_{k,i})$ for $k=1,2$ and $\theta_{2,i} \in \{0,\pi\}$. The energy to minimize becomes 
\begin{equation}
E_{3D} = \frac{1}{2}\sum_{i,j} w_{ij} (\abs{\psi_{1,i} -\psi_{1,j}}^2 + \abs{\psi_{2,i} - \psi_{2,j}}^2 - d_{ij}^2)^2.
\label{E3}
\end{equation}
Below we list  some real-world problems where minimizing $E_{3D}$ can be effectively applied.

\emph{Protein structure determination.}
In protein structure determination~\cite{Protein_Structure}, $\psi_i = (\psi_{1,i}, \psi_{2,i})$ represents the 3D position of amino acid residue $i$ using two complex numbers, $d_{ij}$ is the measured distance between residues (e.g., from nuclear magnetic resonance spectroscopy), and $w_{ij}$ represents the confidence in the measurement, which may depend on factors such as signal strength and the specific experiment used.

\emph{Molecular Conformation.}
For molecular conformation problems~\cite{molecular_design}, $\psi_i$ represents the 3D position of atom $i$, $d_{ij}$ is the known or desired bond length or non-bonded distance between atoms, and $w_{ij}$ might represent the strength of the constraint (e.g., higher for covalent bonds than for non-bonded interactions).

\emph{Earthquake Localization.}
In earthquake localization~\cite{earthquake_location}, $\psi_i$ represents the three-dimensional coordinates of the earthquake epicenter or a seismic station, $d_{ij}$ represents the distance inferred from seismic wave travel times, and $w_{ij}$ might represent the reliability of the time measurement, possibly based on signal clarity or station reliability.

These examples illustrate that distance-based problems are of considerable practical relevance. An efficient solution procedure will advance a wide range of real-world applications.

%\subsection{Solving Distance-Based Problems}

\emph{Solving Distance-Based Problems}. We provide a concise overview of prominent methods for distance‑based problems, using the SNL problem as a representative example, and follow the discussion in Ref.~\cite{Dokmani_REV2015}. In the simplest case of a noiseless SNL problem without anchors, the distance matrix $\D$ forms a Euclidean distance matrix (EDM), which necessarily satisfies the triangle inequality~\cite{strang2019}. The unknown positions can then be recovered via eigenvalue decomposition of the associated Gram matrix—an approach commonly referred to as the matrix method.
Alternatively, one can minimize the energy function~\eqref{E2} (commonly called the s-stress) using gradient descent or its variants. In particular, the alternating descent method has been shown to achieve a success rate of up to $99\%$ in converging to the global minimum~\cite{Parhizkar_thesis}. Another prominent approach relies on semidefinite programming, which exploits the rank properties of the Gram matrix~\cite{BiswasIEEE2006}.
These classical methods generally perform well in the noiseless case but tend to degrade rapidly as noise increases. A detailed comparison of these methods under both noiseless and noisy settings is provided in Ref.~\cite{Dokmani_REV2015}.

More recently, the rapid advancement of machine learning (ML) has led to its application in solving the SNL problem. A comprehensive review and comparison of ML-based methods with traditional optimization approaches can be found in Ref.~\cite{REV_ML_SNL}. Quantitative comparisons between these methods remain challenging due to the diversity of their target applications and design objectives. Reference~\cite{REV_ML_SNL} reviews approximately a dozen different approaches and qualitatively compares their respective advantages and disadvantages.

\section{Proposed Optical Implementation}\label{sec:proposed_optical_implementation}

\emph{Gain-Based Bifurcation (GBB).}
We propose an optical implementation for globally minimizing the energy function in Eqs.~(\ref{E2}) and (\ref{E3}). Our approach uses a network of coupled condensates or lasers~\cite{Claudio_2021}, with the state of the $i$-th condensate described by the wavefunction $\psi_i(t)$. In optical/photonic/coupled light and matter systems, each node of the network evolves according to the dynamical equation~\cite{Lagoudakis_2017,Kalinin_2018,Kalinin2018_SciRep}
\begin{equation}
\dot{\psi_i} = \left[\xi_i(t) - S_i(\psi_i)\right]\psi_i + \sum_{i\neq j} Q_{ij} \psi_j,
\label{eq:osc}
\end{equation}
where $\xi_i(t)$ and $S_i(\psi_i)$ are effective gain and nonlinear saturation of the $i-$th oscillator. The effective gain $\xi_i(t)$ can be the same for all oscillators (uniform pumping) or incorporate a feedback on the amplitude of $\psi_i$. Nonlinear saturation $S_i(\psi_i)$ can take other forms as feedback is implemented~\cite{kalinin2023}.  The network of Eq.~\eqref{eq:osc} serves as the generic platform of the oscillatory network representation. 
%also depend on the amplitude of $\psi_i$ but also on $\psi_j$ with a proper feedback. 

We can adapt the network dynamics of Eq.~\eqref{eq:osc} to minimize $E_{2D}$ or $E_{3D}$ by noticing that the gradient dynamics of $\dot{\psi}_i=-\partial E_{2D}/\partial \psi_i^*$ in two-dimensional space or $\dot{\psi}_{k,i}=-\partial E_{3D}/\partial \psi_{k,i}^*$ in three-dimensional space can be represented by the oscillatory network Eq.~\eqref{eq:osc}.
Indeed, rearranging terms in $\partial E_{2D}/\partial \psi_i^*$, we get
\begin{equation}
 \dot{\psi_i}= (\xi_i - \sum_{j\ne i}S_{ij}) \psi_i + \sum_{j\ne i} Q_{ij}\psi_j, 
 \label{difE2}
\end{equation}
where $\xi_i=\sum_j w_{ij} d_{ij}^2$, $S_{ij}=w_{ij} \abs{ \psi_i - \psi_j }^2$, and $Q_{ij}=w_{ij} \left( \abs{ \psi_i - \psi_j }^2 - d_{ij}^2 \right)$ for the two-dimensional case.
Like any other gradient descent dynamics, this dynamics alone is susceptible to being trapped in local minima of the highly complex $E_{2D}$ energy landscape.
Symmetry-breaking bifurcation analogous to the gain-based principle \cite{Marvin_2023} can be implemented to mitigate this problem by annealing $d_{ij},$ so its value changes from a (randomly chosen) initial value toward the target value over time.
The $\xi_i$, $S_{ij}$, and the $Q_{ij}$ terms in Eq.~\eqref{difE2} are analogous to the pumping term, linear loss term, and the coupling term respectively in a gain-based dissipative system.

 \emph{Canonical Transformation (CT).} All-optical hardware implementation of Eq.~\eqref{difE2} with the feedback required to get the correct $S_{ij}$ and $Q_{ij}$ may be challenging, since it also requires being able to get the resonant terms $\psi_i^*$. We can significantly simplify the required feedback and make it achievable with current technologies by formulating a complementary dual problem, introducing the derivative for $\psi_i$ as an additional set of independent dynamical variables~\cite{gao2017book, xu_arxiv2024}.
We introduce $N \times N$ additional degenerate optical parametric oscillators $\tau_{ij}$ that oscillate with $0$ or $\pi$ phases, so that $\tau_{ij}\in \mathbb{R}$. At equilibrium, we expect
\begin{equation}
 \tau_{ij}=2 \sqrt{w_{ij}}(\abs{\psi_i -\psi_j}^2 - d_{ij}^2). 
 \label{eq:tau_def}
\end{equation}
Using the canonical duality theory, we define the following complementary function
\begin{equation}
    \Phi(\psi_1,\cdots, \psi_N; \tau) = \frac{1}{2}\sum_{i,j}\sqrt{w_{ij}}\tau_{ij}(\abs{\psi_i-\psi_j}^2- d_{ij}^2) - \frac{1}{8}\sum_{i,j}\tau_{ij}^2.
    \label{eq:complimentary}
\end{equation}
The energy \eqref{E2} and the complementary function \eqref{eq:complimentary} have the same stationary points that now can be found in evolving 
\begin{subequations}\label{eq:CT_main}
\begin{align}
 \dot{\tau}_{ij}  & =  \partial_{\tau_{ij}} \Phi  = \sqrt{w_{ij}}(\abs{\psi_i-\psi_j}^2- d_{ij}^2) - \frac{1}{2} \tau_{ij},\label{dtau}\\
 \dot{\psi_i}  & = - \partial_{\psi_{i}^*} \Phi =\Gamma_i \psi_i + \sum_j \sqrt{w_{ij}}\tau_{ij} \psi_j, \label{dpsi}
\end{align}
\end{subequations}
where $\Gamma_i=-\sum_j \sqrt{w_{ij}}\tau_{ij}$ is the feedback on the pump with which $\psi_i$ is populated. The second term in Eq.~\eqref{dpsi} is the usual coupling term with other oscillators in the system with $\sqrt{w_{ij}}\tau_{ij}$ being the strength of these couplings. Equation (\ref{dtau}) evolves $\tau_{ij}$ using the resonant excitation that depends on the states of $\psi_i$ and $\psi_j$ and the constant dissipation with rate $1/2$. Again, $d_{ij}(t)$ can be annealed over time to the expected value. Both equations are straightforward to implement optically but require electronic feedback.

\section{Numerical Results}\label{sec:numerical_result}

\begin{figure}[t]
  \centering
  \includegraphics[width=\linewidth]{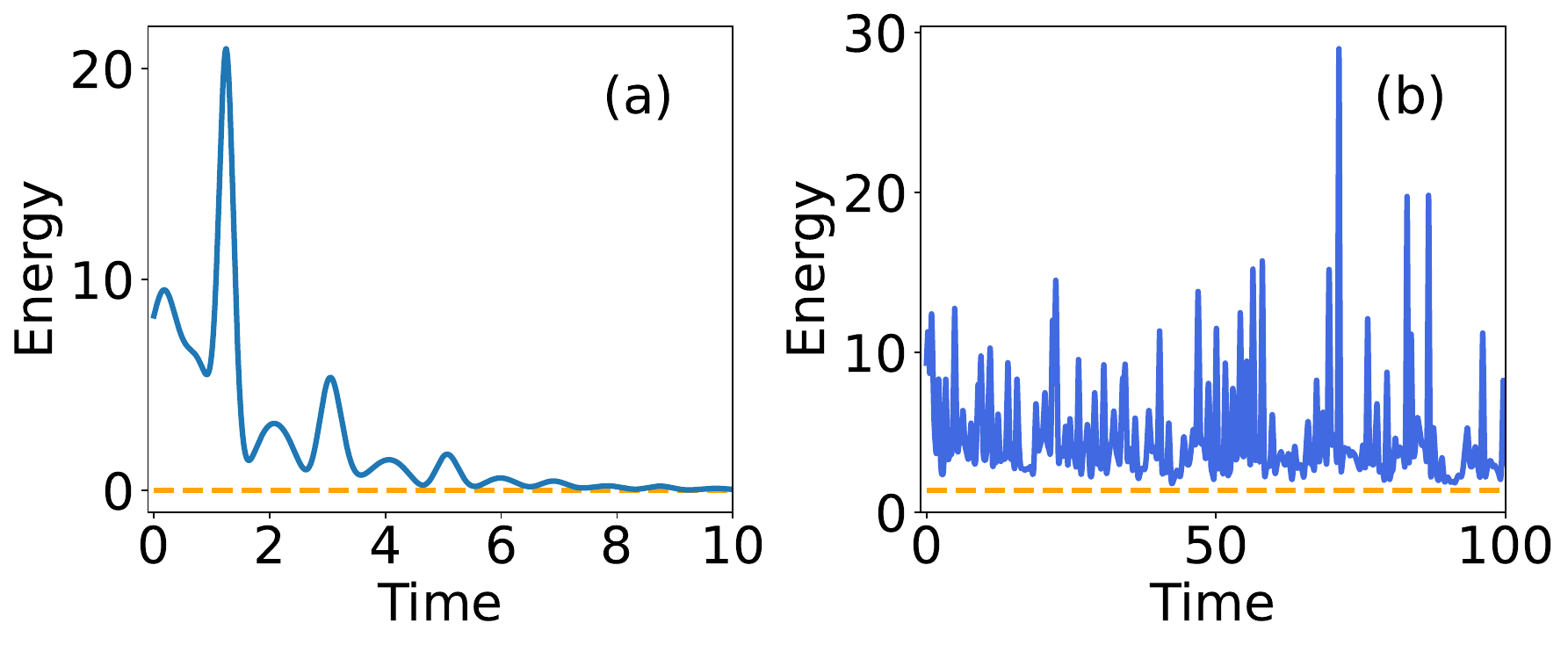}
  \caption{Energy evolution in the numerical integration of Eqs. (\ref{dtau}-\ref{dpsi}) for   $N=10$ SNL problem without anchors, with all weights $w_{ij}=1$.
(a) The $\D$ matrix is generated from a random distribution of points in a unit-length box, making it a valid Euclidean distance matrix (EDM).
(b) The $\D$ matrix is constructed from random values sampled uniformly from the interval $(0, 1)$ and does not correspond to any point configuration in the two-dimensional Euclidean space; that is, it is a non-EDM matrix.
The initial condition sets $\{\psi_i\}$ randomly within a square box of side length given by the maximum element of the $\D$ matrix. The initial $\{\tau_{ij}\}$ is sampled uniformly from the interval $(-1, 1)$.
The dashed orange lines indicate the ground state energy computed using SciPy's global optimization solver. The ground state energy is zero for EDM cases, whereas for non-EDM cases it takes a finite positive value.} %\nb{Specify the initial condition for the simulation here and in the other captions.}}
  \label{fig:eng_simple}
\end{figure}

\subsection{Synchronous and Asynchronous Update}

In this subsection, we first demonstrate the effectiveness of the CT method, and then discuss its limitations and corresponding remedies.  
We begin by solving Eq.~\eqref{eq:CT_main} using the fixed-step fourth-order Runge-Kutta method (RK4), with results presented in Fig.~\ref{fig:eng_simple}. If $\D$ is an EDM, then starting from randomly distributed initial values, the system evolves to its ground state energy (in this case, zero), as claimed in Ref.~\cite{gao2017book}. We have verified this behavior for larger values of $N$, and this global convergence consistently occurs across all tested instances. This is unsurprising, as the current noiseless scenario presents a relatively simple problem, for which many methods can achieve an almost $100\%$ success rate within reasonable tolerance~\cite{Dokmani_REV2015}. However, complications arise in the presence of noise, when the $\D$ matrix becomes a non-EDM.

Figure~\ref{fig:eng_simple}(b) illustrates an example of energy evolution for a non-EDM case. As observed, the system fluctuates chaotically and fails to settle into a stationary state. This behavior can be understood as follows: the $\tau_{ij}$ values, which are expected to vanish near the global minimum, can never be exactly zero because it is impossible to find a set of ${\psi_i}$ that cancels every element of the $\D$ matrix [see Eq.~\eqref{eq:tau_def}]. These persistent non-zero values act as forces that continually drive the points to evolve, even if they are momentarily near a local or global minimum configuration.

To improve the situation, we observe from Fig.~\ref{fig:eng_simple}(b) that the fluctuations consist of a high-frequency oscillation superimposed on a lower-frequency oscillation. This suggests that the two dynamical equations in Eq.~\eqref{eq:CT_main} may be evolving on different time scales, preventing them from synchronizing and reaching a stationary state. 
Intuitively, during the gradient ascent for $\tau_{ij}$ and the gradient descent for $\psi_i$, each variable influences the energy landscape of the other. It is therefore plausible that one of them, say $\tau_{ij}$, does not have sufficient time to settle into a nearby local minimum before $\psi_i$ updates and alters the landscape. This landscape shift displaces the previous local minimum, forcing $\tau_{ij}$ to adapt to a new direction. As a result, the dynamics may continue fluctuating indefinitely without converging to a stationary state. Such two-frequency evolution is a general phenomenon and can also be observed in the EDM case shown in Fig.~\ref{fig:eng_simple}(a), where the fluctuation amplitudes of both frequencies gradually decrease and eventually converge to the global minimum.

Based on the above physical interpretation, we propose an asynchronous update scheme for Eq.~\eqref{eq:CT_main} using a fixed-step RK4 solver. 
Specifically, we evolve the dynamics of $\tau_{ij}$ sequentially with a base time step $\Delta t$, while keeping $\psi_i$ fixed. After a delay of $\Delta t_{\psi}$, we then update $\psi_i$ by $\Delta t$. (Numerically, we find that introducing a delayed update, either for $\tau$ or $\psi$, can enhance convergence; however, in the case of delayed $\tau$, the convergence is less prominent.) The chosen approach is particularly advantageous because $\tau_{ij} \in \mathbb{R}$, making it easier to implement and update than the complex-valued $\psi_i$. This is especially relevant for physical platforms such as the coherent Ising machine, which naturally supports real-valued encoding.

\begin{figure}[t]
  \centering
  \includegraphics[width=\linewidth]{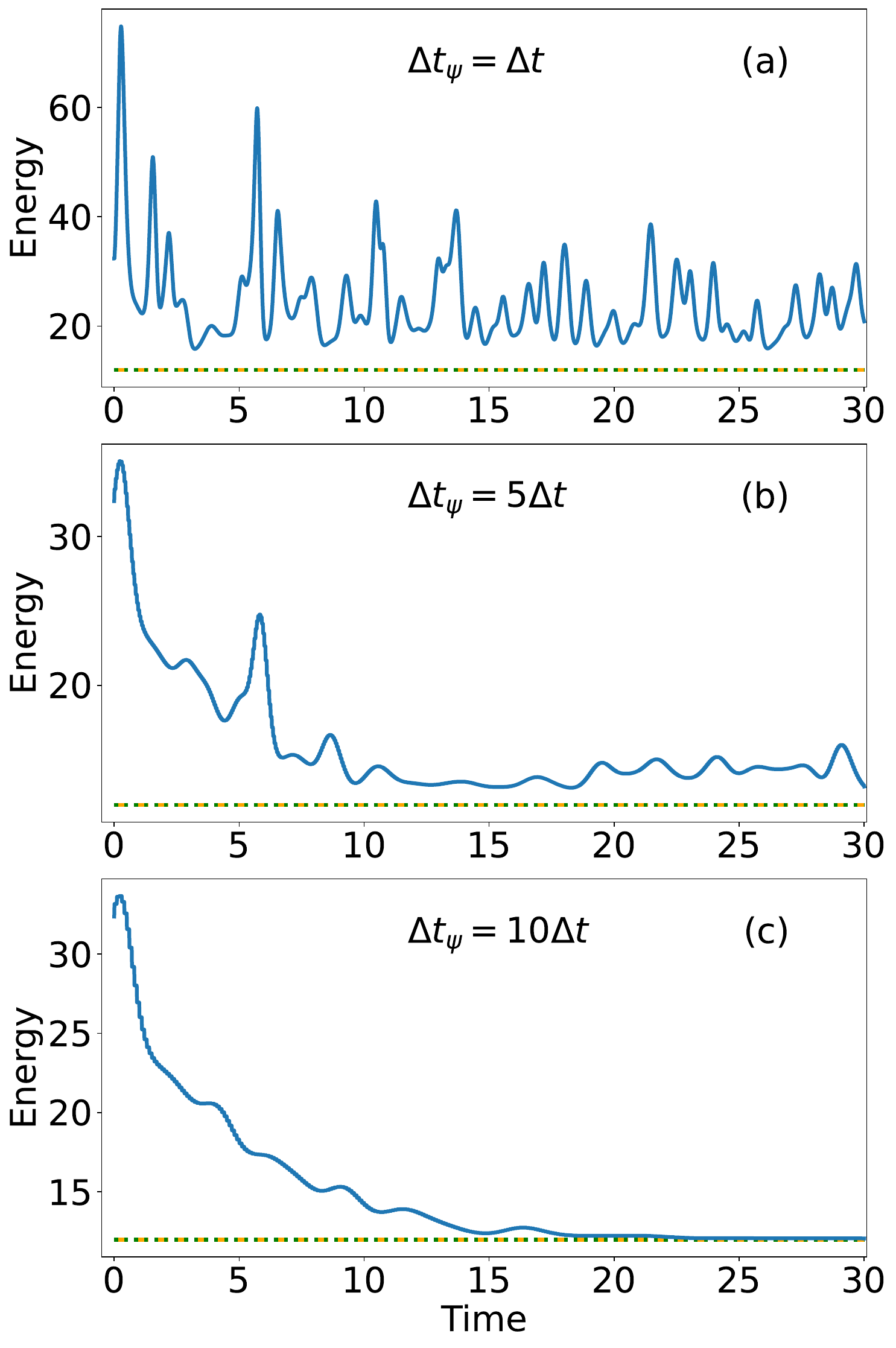}
  \caption{Energy evolution in the asynchronous update of a $N=20$ two-dimensional non-EDM SNL problem without anchors, using uniform weights $w_{ij} = 1$. The $\D$ matrix is
  constructed from random values sampled uniformly from the
  interval $(0,1)$. The system is solved using an RK4 integrator with a base time step $\Delta t = 0.01$, and all runs start from the same initial configuration. The initial condition sets $\{\psi_i\}$ randomly within a square box of side length given by the maximum element of the $\D$ matrix. The initial $\{\tau_{ij}\}$ is sampled uniformly from the interval $(-1, 1)$. Panels show different update intervals for $\psi$: (a) $\Delta t_{\psi} = \Delta t$, (b) $\Delta t_{\psi} = 5\Delta t$, and (c) $\Delta t_{\psi} = 10\Delta t$. Orange dashed lines represent the ground state energy found using SciPy's global optimizer, while green dashed lines indicate the best energy obtained from SciPy's local optimizer over $200$ random initializations.
}
  \label{fig:eng_delay}
\end{figure}

Figure~\ref{fig:eng_delay} demonstrates the advantage of asynchronous updates for a two-dimensional non-EDM SNL problem with $N=20$. As anticipated, the synchronous update using the RK4 solver results in a fluctuating state. However, by progressively increasing the update delay for $\psi_i$, these rapid fluctuations are smoothed out. Ultimately, with a sufficiently large $\Delta t_{\psi}$, the system converges to the global energy minimum within a reasonable tolerance of less than $1\%$ within the given simulation time. Figure~\ref{fig:eng_delay}(b,c) clearly shows the stepped behavior of the energy function. Since Eq.~\eqref{E2} depends solely on $\psi_i$ and not on $\tau_{ij}$, this discretization effectively allows $\tau_{ij}$ more time to relax, thereby suppressing rapid fluctuations.

%We now return to the asynchronous update dynamics. 
Increasing the value of $\Delta t_{\psi}$ can lead to a smoother energy evolution curve. However, this also slows down convergence to the ground state, as asynchronous updates inherently decelerate the system’s dynamics. In practice, due to computational time constraints, it may be necessary to employ advanced search strategies such as binary search~\cite{clrs2009introduction} to identify the optimal $\Delta t_{\psi}$ that satisfies the desired tolerance. It is worth noting that the asynchronous update scheme can also be applied to the anchored case, together with weights that are not all equal to one. The resulting behavior is qualitatively similar to that shown in Fig.~\ref{fig:eng_delay}. However, in such complex scenarios, a larger $\Delta t_{\psi}$ may be required, which leads to a significant slowdown in the system's dynamics. In the next subsection, we propose an alternative method to overcome this limitation.

\subsection{Steepened Gradient Dynamics}\label{subsec:Steepened_gradient}

From the above discussion, we know that the coupled dynamical system \eqref{eq:CT_main} exhibits two different time scales. The asynchronous update method aims to slow down the rapidly evolving component in order to enhance convergence to a stationary state. Alternatively, we can accelerate the slowly evolving component to achieve a similar effect. To proceed, we multiply a positive constant $\eta_\tau$ to the right-hand side of Eq.~\eqref{dtau}, effectively steepening its gradient. This gives
\begin{subequations}\label{eq:CT_eta}
\begin{align}
 \dot{\tau}_{ij} &= \eta_\tau \left[ \sqrt{w_{ij}}(|\psi_i-\psi_j|^2- d_{ij}^2) - \frac{1}{2} \tau_{ij}\right],\label{dtau_eta}\\
 \dot{\psi_i} &=\Gamma_i \psi_i + \sum_j \sqrt{w_{ij}}\tau_{ij} \psi_j, \label{dpsi_eta}
\end{align}
\end{subequations}
where $\Gamma_i$ is defined the same as in Eq.~\eqref{dpsi}. Clearly, Eqs.(\ref{dtau_eta} - \ref{dpsi_eta}) and Eqs.(\ref{dtau} - \ref{dpsi})have the same fixed points.  We can now solve Eq.~\eqref{eq:CT_eta} using a synchronous update method and expect improved convergence. Specifically, we employ the adaptive-step RK4 solver provided by SciPy, which can significantly reduce the total number of integration steps compared to our previous fixed-step RK4 asynchronous solver.

\begin{figure}
  \centering
  \includegraphics[width=\linewidth]{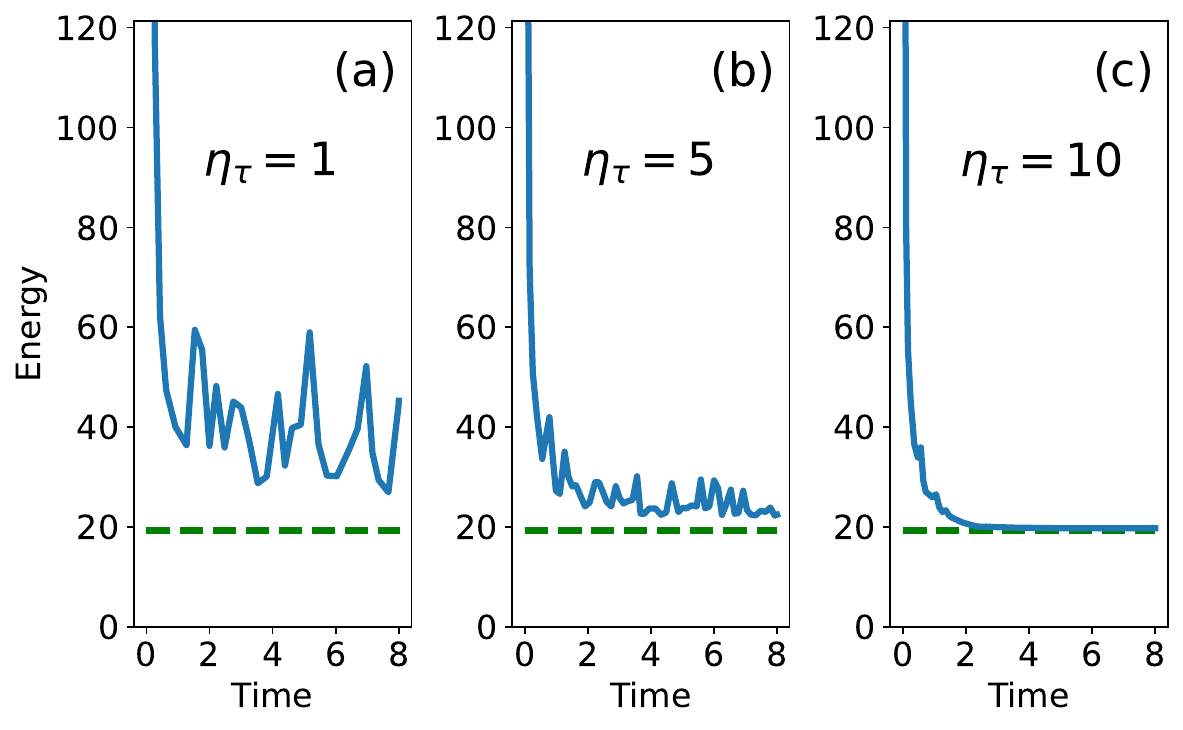}
  \caption{Energy evolution of the synchronous adaptive step updates for a two-dimensional non-EDM SNL problem with $N=35$ nodes and 10 anchors that are sampled from a unit box. The $\D$ matrix element is sampled uniformly from the interval $(0,1)$, then adjusted to accommodate the anchor distances.
  The weights $w_{ij}$ are set to $1$ for all anchors, and randomly chosen from $\{0, 1\}$ with $50\%$ probability for the free points. All runs begin from the same initial configuration.
  The initial condition sets $\{\psi_i\}$ randomly within a square box of side length given by the
  sum of the maximum element of the $\D$ matrix and the maximum anchor amplitude. The initial $\{\tau_{ij}\}$ is sampled uniformly from the interval $(-1, 1)$.
  Panels illustrate the effect of different gradient amplification factors: (a) $\eta_\tau = 1$. (b) $\eta_\tau = 5$. (c) $\eta_\tau = 10$. Green dashed lines indicate the best energy found by SciPy’s local optimizer over $200$ random initializations.
}
  \label{fig:eng_eta}
\end{figure}

Figure~\ref{fig:eng_eta} shows three sets of energy evolution curves, all starting from the same initial conditions but with different values of $\eta_\tau$. The relative and absolute tolerances for the adaptive steps are set to $10^{-3}$ and $10^{-6}$, respectively. For $\eta_\tau = 1$, the system does not settle into a stationary state, as expected. As $\eta_\tau$ increases, both the amplitude and frequency of the fluctuations decrease. When $\eta_\tau = 10$, the system converges rapidly to the ground state energy, reaching a stationary value by simulation time $T = 2.5$ (approximately 30 steps), with energy within $2\%$ of the ground state obtained via brute-force optimization. We have changed the plotting style of the steepened gradient method to visually distinguish it from the asynchronous update method discussed in the previous subsection.

We note that the problem presented in Fig.~\ref{fig:eng_eta} represents the most general case of the two-dimensional SNL problem. In such cases, the performance of SciPy's global optimization solvers deteriorates significantly, consuming more time but giving poor results. Therefore, in this and subsequent subsections, we consider the energy obtained via the brute-force method as the ground state energy for general problem instances.

As the SNL problem becomes more challenging, the performance of the CT method also deteriorates. To enhance its success rate, we introduce annealing techniques, which will add an additional time scale to the system's dynamics. The steepened gradient method is well-suited to handle this complexity, as it allows for straightforward adjustment of $\eta_\tau$ to accommodate the new time scale without significantly slowing down the simulation.

\begin{figure}
  \centering
  \includegraphics[width=\linewidth]{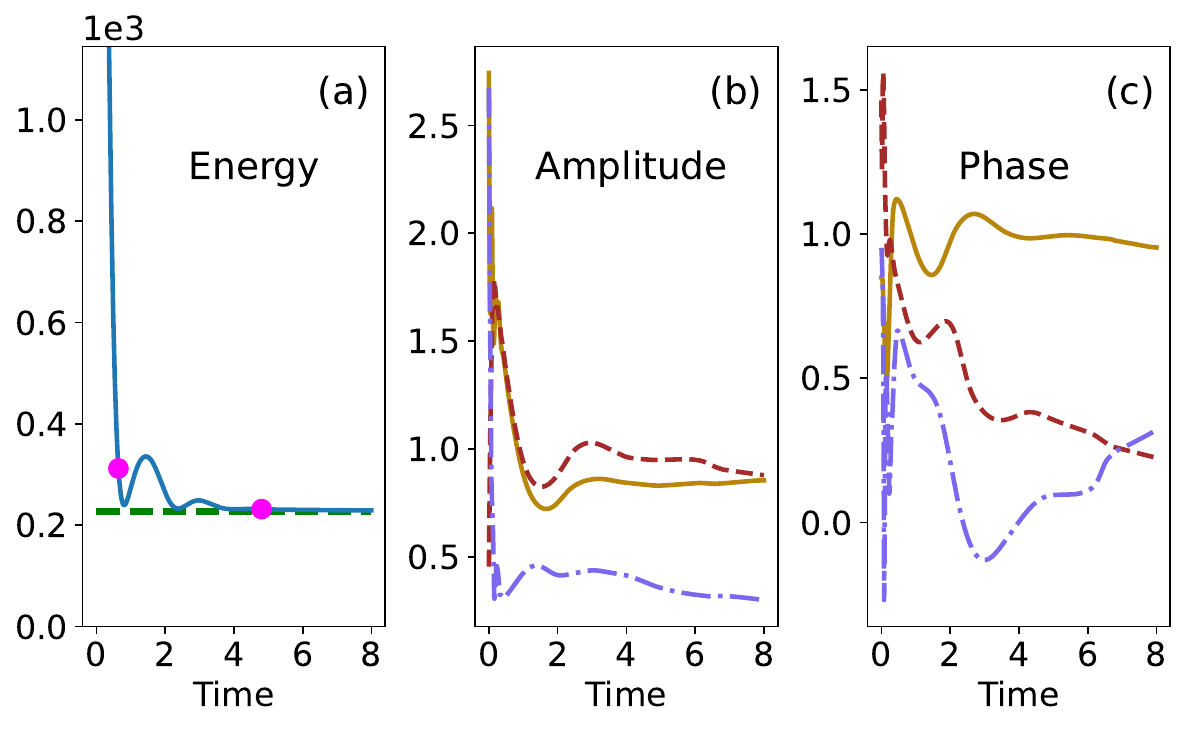}
  \caption{Simulation of a two-dimensional non-EDM SNL problem with $N=100$ points and $20$ anchors that are sampled from a unit box. The $\D$ matrix element is sampled uniformly from the interval $(0,1)$, then adjusted to accommodate the anchor distances.
  Weights $w_{ij}$ are set to $1$ for all anchors, and randomly chosen from $\{0, 1\}$ with $50\%$ probability for free points. The initial condition sets $\{\psi_i\}$ randomly within a square box of side length given by the sum of the maximum element of the $\D$ matrix and the maximum anchor amplitude. The initial $\{\tau_{ij}\}$ is sampled uniformly from the interval $(-1, 1)$.
 (a) Energy evolution (blue) of the synchronous adaptive-step updates with $\eta_\tau = 100$ and annealing. The green dashed line shows the ground state energy obtained via SciPy’s local optimizer over $200$ random initializations. Magenta dots indicate the time points corresponding to the evolution snapshots in Fig.~\ref{fig:snapshot}.  
(b) Amplitude and (c) phase evolution of three selected free points.
}
  \label{fig:eng_evo}
\end{figure}

Figure~\ref{fig:eng_evo} shows an example simulation of steepened gradient method with annealing. The annealing scheme is given by $d_{ij}\to d_{ij}\left[1+c_1 \sin(\omega t+c_2)\exp(-c_3 t)\right]$, where $c_{1,2,3}$ are constants and $\omega$ is the oscillation frequency.
Figure~\ref{fig:eng_evo}(a) shows the energy (blue) near the ground state energy (green dashed), and the annealing oscillation can be clearly seen for $t<5$. In this example, annealing enables the dynamical system to reach a state with energy within $0.7\%$ of the ground state.

Figures~\ref{fig:eng_evo}(b)(c) show the amplitude and phase evolution for three selected free points. It is apparent that the system has not yet reached a stationary state, as both the amplitude and phase of the selected point have not saturated. We expect that extending the simulation time would lead to a lower-energy configuration. However, since further improvements are likely to be marginal, it is important to balance computational cost and accuracy by choosing an appropriate tolerance.

%%energy evolution for CT

\begin{figure}[ht]
  \centering
  \includegraphics[width=\linewidth]{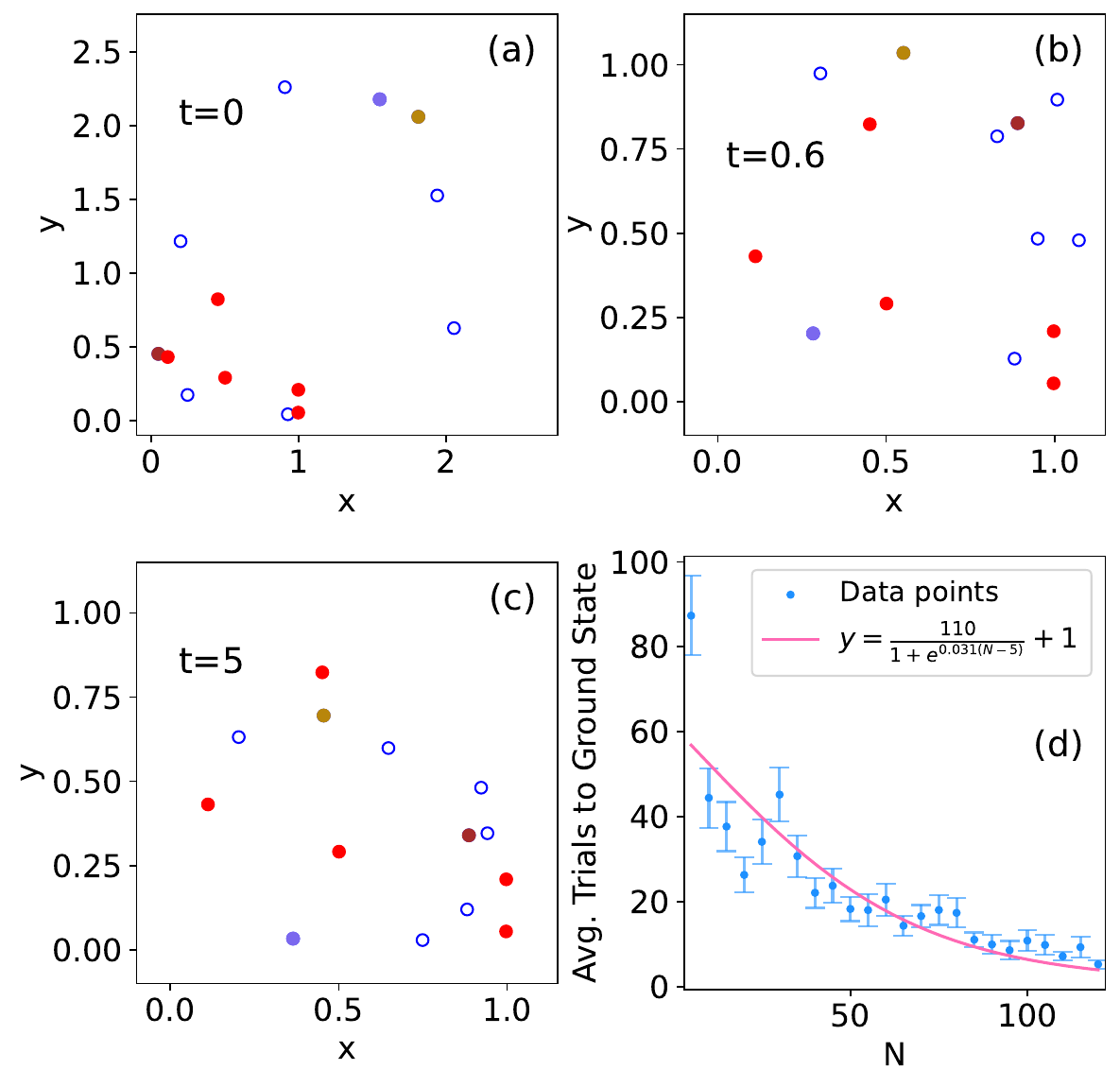}
  \caption{(a)-(c) Snapshot of the spatial evolution of selected anchored points (solid red) and free points (empty blue) for the simulation in Fig.~\ref{fig:eng_evo}. The three solid-color points correspond to data in Fig.~\ref{fig:eng_evo}(b)(c). (d) The average number of random trials needed to reach the ground state energy for a problem of size $N$ with a fixed anchor-to-$N$ ratio of $\lfloor 0.1N \rceil$. Parameters are the same as in Fig.~\ref{fig:eng_evo}. The pink curve gives a fit to the logistic function that saturates to one: $y=a/[1+e^{b(N-c)}]+1$. The best-fit parameters, along
with their standard deviations, are: $a=110\pm120$, $b=0.031\pm0.014$, and $c=5\pm65$.  The decaying behavior indicates a potential issue in the normalized problem for large $N$ (see text for details).
}
  \label{fig:snapshot}
\end{figure}

Figures~\ref{fig:snapshot}(a)--(c) show snapshots of the real-space distribution for selected anchor and free points. The simulation starts from random initial positions distributed within a box, whose length scale is determined by the sum of the maximum element of the $\D$ matrix and the maximum anchor amplitude. 
The figures illustrate how the selected free points evolve toward the anchor region and oscillate before settling. The solid-colored points correspond to the time evolution data shown in Fig.~\ref{fig:eng_evo}(b)(c). The gold and brown points eventually reach similar amplitudes, but they correspond to different locations due to their differing phases.  
Under the annealing parameters: $c_1 = 1.2$, $c_2 = \pi/2$, $c_3 = 0.75$, and $\omega = 2$, the simulation completes in approximately $200$ steps. In general, increasing either $c_3$ or $\eta_\tau$ leads to a faster convergence to a stationary state.

%%evolution snapshot

Figure~\ref{fig:snapshot}(d) shows the average number of trials required for the steepened gradient dynamics, starting from random initial states, to reach a ground state within a tolerance of $0.5\%$. For each value of $N$, we fix the anchor-to-$N$ ratio at $10\%$, taken as the nearest integer of $\lfloor 0.1N \rceil$. 
For each $N$, we generate $100$ random problem instances, each defined by a non-EDM $\D$ matrix and a random $\W$ matrix. For each instance, the simulation is run from $200$ different random initial states until the ground state energy is reached within the specified tolerance. (If none of the initial states lead to a satisfactory solution, the trial count for that instance is set to $200$.) To keep the simulation converging across $N$, we have set $\eta_\tau=50$.

We expected the average number of trials to grow exponentially with $N$, due to the increasing complexity of the search space. However, the simulation data shows a different pattern: the average number of trials initially fluctuates for small $N$, but then quickly drops as $N$ increases. 
This behavior can be understood by considering the energy landscape of the SNL problem. As $N$ increases, the landscape becomes densely populated with shallow local minima. 
This is because $d_{ij} \in (0,1)$, and by operating on small values, the energy function in Eq.~\eqref{E2} reduces the difference between local and global minima.
As a result, it becomes difficult to distinguish a local minimum from the global one solely judged by the energy value. This suggests, as $N$ increases, attention should be paid to ensure the system is appropriately scaled. Although normalizing the distance matrix is a common practice to improve numerical stability, it is also important to examine the problem in its original scale, since the energy function is not scaling linearly.

%as i reduces the disparity between the largest and smallest eigenvalues of the distance matrix, thereby mitigating issues related to ill-conditioning. In the following discussions, we will demonstrate the effectiveness of both the canonical transformation method and the gain-based method without relying on normalization.

\subsection{Integrating Asynchronous Updates with Steepened Gradients}

\begin{figure}[t]
  \centering
  \includegraphics[width=0.9\linewidth]{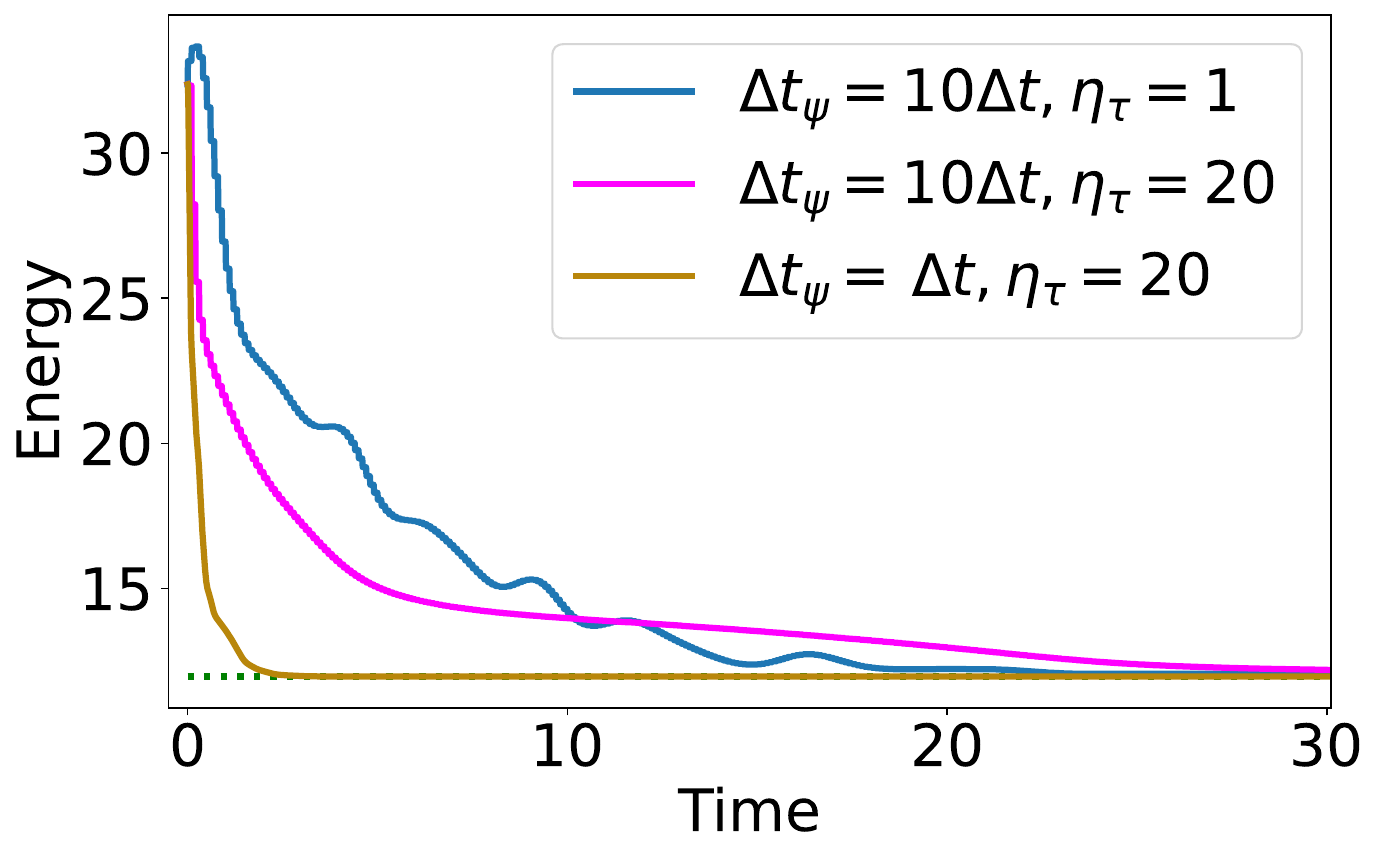}
  \caption{Comparison of energy evolution under the asynchronous update scheme with $\Delta t = 0.01$, for various values of $\Delta t_\psi$ and $\eta_\tau$, in a two-dimensional non-EDM SNL problem with $N = 20$, no anchors, uniform weights $w_{ij} = 1$, and $d_{ij} \in (0,1)$.
  The initial condition sets $\{\psi_i\}$ randomly within a square box of side length given by the maximum element of the $\D$ matrix. The initial $\{\tau_{ij}\}$ is sampled uniformly from the interval $(-1, 1)$.
  }
  \label{fig:delay_and_eta}
\end{figure}

We have presented two techniques to enhance the CT method for solving non-EDM SNL problems. The choice of an appropriate technique for optical implementation should depend on the characteristics of the specific optical platform, particularly whether feedback is computationally expensive. Figure~\ref{fig:delay_and_eta} shows simulations of synchronous and asynchronous updates, both with and without the steepened gradient technique. All three approaches effectively solve the non-EDM SNL problem. In particular, the quantized structure observed in the energy evolution curves suggests that, for optical implementations involving feedback, the asynchronous update technique may offer advantages by reducing the total number of feedback operations.

However, asynchronous updates rely on fixed-step updating rules, which may not be feasible in analog optical systems. In such cases, the steepened gradient technique can be more advantageous, as it not only allows tuning of the dominant time scale to enhance convergence, but also accommodates annealing strategies that introduce additional time scales into the dynamics.

Lastly, combining both techniques does not appear to yield additional improvement. This is because the primary challenge in the CT method stems from the mismatch between two distinct time scales, and either technique alone is sufficient to effectively mitigate this issue. As a result, the magenta curve in Fig.~\ref{fig:delay_and_eta} does not outperform either the blue or the gold curve.

\subsection{Comparison With the Gain-Based Bifurcation Method}

We now proceed to analyze the success rate statistics. We fix the dimension of the $\D$ matrix to $N\times N=35\times 35$ and set the number of anchored points to $10$. Then we repeat the simulation by randomly generating $100$ distinct sets of $\D$ and $\W$, where $d_{ij} \in (0,10)$ and $w_{ij} \in \{0, 1\}$ for free points, with each set simulated $200$ times under random initial conditions. We use the SciPy adaptive RK4 solver for all methods. 
For the CT steepened gradient method, we set $\eta_\tau=10^3$.

\begin{figure}[t]
  \centering
  \includegraphics[width=\linewidth]{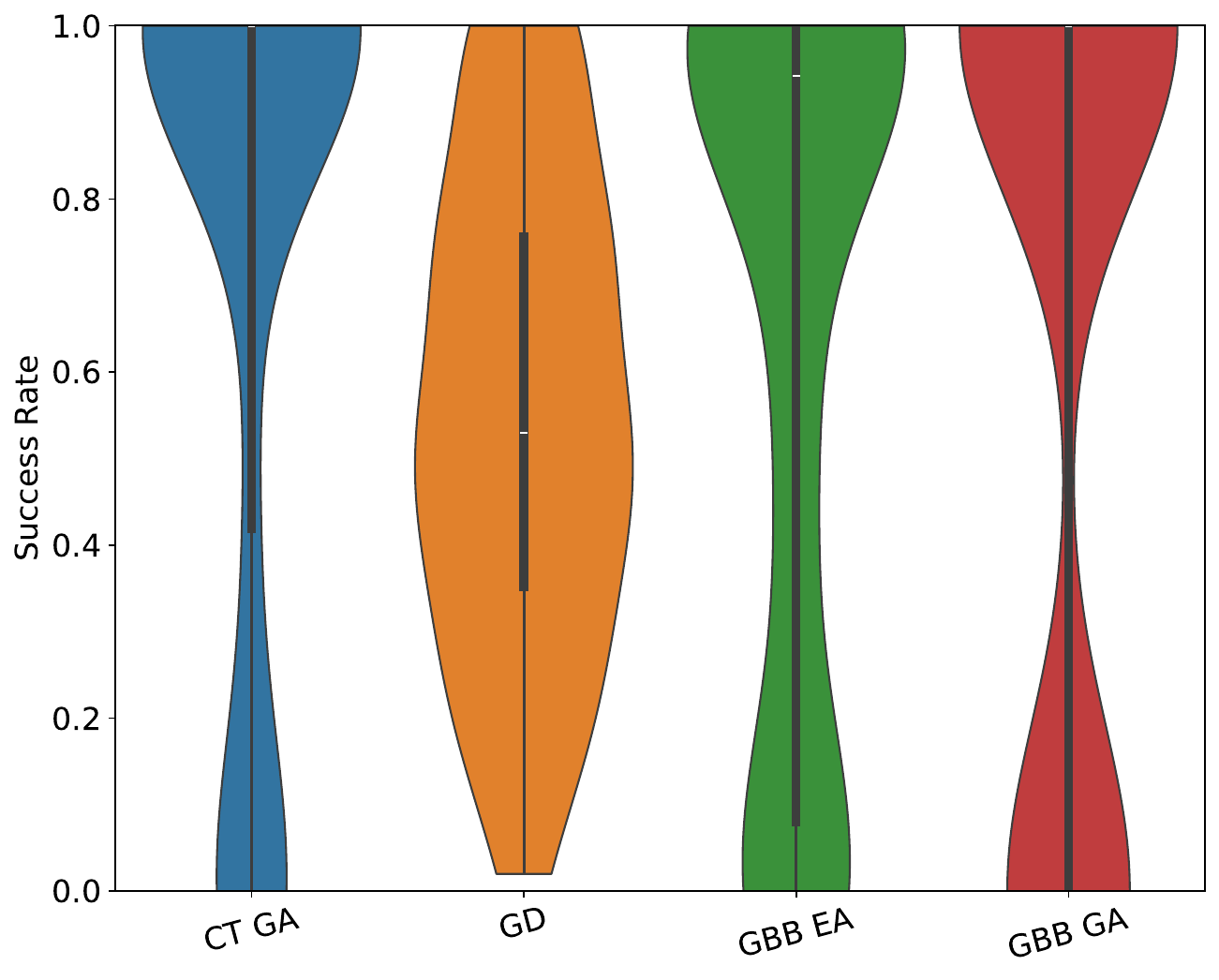}
  \caption{Success rate comparison for a two-dimensional non-EDM SNL problem with $N = 35$ and $10$ anchors, using randomly sampled \( w_{ij} \in \{0, 1\} \), where each value is chosen with equal probability for free points and $w_{ij}=1$ for anchors. The $\D$ matrix has $d_{ij} \in (0, 10)$.
  There are $100$ distinct sets of $\D$ and $\W$ matrices, with each set repeated $200$ times for random initial states. The initial condition sets $\{\psi_i\}$ randomly within a square box of side length given by the sum of the maximum element of the $\D$ matrix and the maximum anchor amplitude. The initial $\{\tau_{ij}\}$ is sampled uniformly from the interval $(-1, 1)$. 
  The tolerance for success is below $10^{-3} E_g$, where $E_g$ is determined as the minimum result obtained from SciPy’s local optimizer over $100$ random initializations. The symbols representing different annealing schemes are as follows: CT GA -- canonical transformation steepened gradient with global annealing, GD -- gradient descent, GBB EA – gain-based bifurcation method with element-wise annealing, and GBB GA -- gain-based bifurcation method with global annealing.
  }
  \label{fig:suc_rate_compare}
\end{figure}

Figure~\ref{fig:suc_rate_compare} presents the violin plot that compares success rates for various methods.
The classical gradient descent method performs poorly, achieving a median success rate of only around $50\%$. In contrast, both the CT and GBB methods reach a median success rate of $100\%$ when equipped with suitable annealing schemes. 

For the GBB method, different annealing schemes were implemented to vary the $\D$ matrix components in the pumping term [i.e. the $\xi_i$ term in Eq.~\eqref{difE2}], and they are described by the following expressions
\begin{equation}
d_{ij}(t) =
\begin{cases}
d_{ij}^\prime, & \text{$t > t_{anneal}$} \\
d_{ij}^\prime + d_{ij}^\prime r_{ij}P(t)\cos{(\omega_{ij} t)}, & \text{$t \le t_{anneal}$}
\end{cases}
\end{equation}
where $P(t)$ is a strictly decreasing function that equals $0$ for all $t \ge t_{anneal}$, $r_{ij}$ are random positive real number that determine the initial oscillatory amplitude of each element, $\omega_{ij}$ are angular frequency of oscillation, and $d_{ij}^\prime$ are the $\D$ matrix elements given by the SNL problem instance.
Note that $r_{ij}$ and $\omega_{ij}$ must be symmetric to ensure $d_{ij}$ remains symmetric at all times.
In our simulations, $P(t)$ always decreases linearly from $1$ to $0$, and $\omega_{ij} = 2\pi$.
When all initial oscillatory amplitudes were set to the same value $r_{ij} = r$, this annealing scheme is denoted as the global annealing (GA) scheme in Fig.~\ref{fig:suc_rate_compare}.
When the initial oscillatory amplitudes $r_{ij}$ were uniformly randomly distributed in the range $(0.1, 2.0)$, this annealing scheme is referred to as the element-wise annealing (EA) scheme in Fig.~\ref{fig:suc_rate_compare}.

%%success rate compare

As shown in Fig.~\ref{fig:suc_rate_compare}, the GBB method, with either of the annealing scheme, significantly enhances the median success rate. However, this comes at the cost of a substantial fraction of problem cases with zero success probability, as indicated by the middle box plot of the red distribution extending to zero.
In contrast, the CT method offers distinct advantages, achieving a higher median success rate while also reducing the total number of failures.

\begin{figure}
    \centering
    \includegraphics[width=\linewidth]{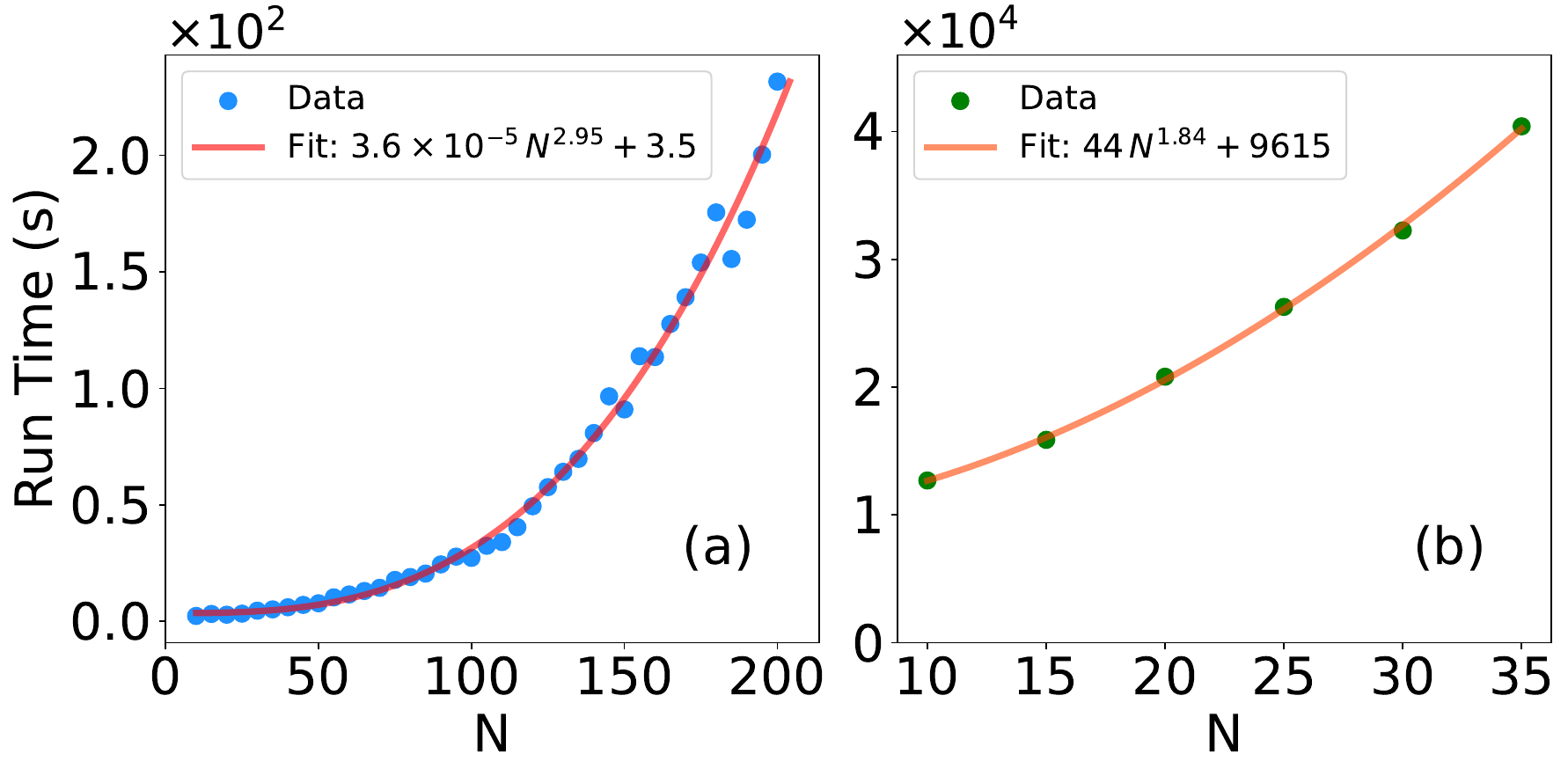}
    \caption{
     Runtime comparison of the CT steepened gradient method as a function of problem size $N$, with the anchor-to-$N$ ratio fixed at $0.2$; that is, the number of anchors is set to the nearest integer of $\lfloor 0.2N \rceil$. Data are fitted to the expression: $y=a N^b+c$.
     (a) Runtime for a single run at a given $N$, using $\eta_\tau = 5.5 \times 10^3$. The best-fit parameters, along with their standard deviations, are: $a_s=(3.6\pm2.0)\times10^{-5}$, $b_s=2.946\pm0.108$, and $c_s=3.5\pm1.8$.
     (b) Runtime for a batch run at a given $N$, consisting of $100$ independent problem instances, each repeated with $200$ different initial conditions, using $\eta_\tau = 1.3 \times 10^3$. 
     The best-fit parameters are: $a_b=44\pm16$, $b_b=1.839\pm0.095$, and $c_b=9615\pm657$. 
     All other simulation parameters are the same as those used in Fig.~\ref{fig:suc_rate_compare}. 
     Runtime data were collected on the Cambridge CSD3 IceLake system with one task assigned to each CPU core for every value of $N$ to avoid CPU number dependence and to exclude parallel overhead from the data.}
    \label{fig:runtime}
\end{figure}

Note that in Fig.~\ref{fig:suc_rate_compare} we have increased the numerical range of the elements in the $\D$ matrix. As discussed in Sec.~\ref{subsec:Steepened_gradient}, normalizing the maximum element of $\D$ to one compresses the energy landscape: the depths of most local minima become nearly indistinguishable from that of the global minimum. As a result, classical gradient-based methods may appear to converge successfully on the normalized problem, but once the solution is rescaled back, it may in fact lie far from the true global minimizer. We have observed that as the sampling range of $d_{ij}$ increases, for instance to $(0,10)$, the depth variation among minima grows, and the success rate of methods like gradient descent correspondingly deteriorates quickly.

On the other hand, increasing the numerical range of the $\D$ matrix elements significantly increases the computational time required for simulation. This is because increasing the range directly amplifies the term in \eqref{dtau} by a quadratic factor, which in turn inflates the eigenvalues of the Jacobian and renders the system stiff. Although an adaptive integrator can compensate by reducing its step size, the combination of the number of state variables and a minimal allowable step size causes the total number of steps to grow super-quadratically.

Figure~\ref{fig:runtime} shows the simulation runtime of the CT steepened gradient method for different values of $N$, using the same parameters as in Fig.~\ref{fig:suc_rate_compare}, with a suitable $\eta_\tau$ selected based on the largest value of $N$.
While a single simulation with $d_{ij} \in (0,10)$ can handle $N = 200$ in approximately 3 minutes, collecting statistics over 100 problems—each with 200 random initial states—at each value of $N$ becomes vastly more time-consuming. In practice, $N = 35$ was the largest system size for which we were able to obtain statistically significant success-rate data within the constraints of our allocated HPC resources.
The implementation of physics-based networks enables massive parallelism and continuous-time integration on dedicated hardware, offering orders-of-magnitude speed-ups over software-based solvers and thereby allowing both single-trial and large-batch evaluations at much larger $N$ without prohibitive runtimes.

\section{Conclusions} \label{sec:conclusions}

We propose a novel approach to solving distance-based optimization problems using optical hardware. The underlying framework can be formulated through either the Canonical Transformation (CT) method or the Gain-Based Bifurcation (GBB) method. 
To enhance the CT method, we further introduce two numerical techniques: asynchronous update and steepened gradient. Numerical simulations on two-dimensional wireless sensor network localization problems demonstrate that both the CT and GBB methods can efficiently solve the general problem.
Since both methods naturally operate in the complex domain, they are particularly well suited for implementation on modern optical systems.

For three-dimensional problems, the dynamical equations can be derived analogously from the energy function in Eq.~\eqref{E3}, enabling the application of the canonical transformation in a similar manner. We anticipate that both the CT and GBB methods will achieve higher success rates than the alternating descent method across the range of distance-based problems discussed in Sec.~\ref{sec:distance_based}, owing to their noise tolerance capability. Therefore, our methods provide efficient solutions to important challenges in science and engineering.

Future work will explore experimental implementations of these methods and benchmark their performance such as accuracy, speed, and energy efficiency, against digital computational algorithms. We also aim to extend the analysis to more complex scenarios, including cases where objects are constrained to curved surfaces or reside within non-simply connected domains.

\begin{acknowledgments}
We thank Hayder Salman for insightful discussions. We acknowledge the support of the Cambridge Service for Data-Driven Discovery (CSD3) for providing high-performance computing resources. This work was supported by HORIZON EIC-2022-PATHFINDERCHALLENGES-01 HEISINGBERG Project 101114978, Weizmann-UK Make Connection Grant 142568, and  the EPSRC UK Multidisciplinary Centre for Neuromorphic Computing award UKRI982.
\end{acknowledgments}

\bibliography{SNL_reference}

\begin{thebibliography}{48}%
\makeatletter
\providecommand \@ifxundefined [1]{%
 \@ifx{#1\undefined}
}%
\providecommand \@ifnum [1]{%
 \ifnum #1\expandafter \@firstoftwo
 \else \expandafter \@secondoftwo
 \fi
}%
\providecommand \@ifx [1]{%
 \ifx #1\expandafter \@firstoftwo
 \else \expandafter \@secondoftwo
 \fi
}%
\providecommand \natexlab [1]{#1}%
\providecommand \enquote  [1]{``#1''}%
\providecommand \bibnamefont  [1]{#1}%
\providecommand \bibfnamefont [1]{#1}%
\providecommand \citenamefont [1]{#1}%
\providecommand \href@noop [0]{\@secondoftwo}%
\providecommand \href [0]{\begingroup \@sanitize@url \@href}%
\providecommand \@href[1]{\@@startlink{#1}\@@href}%
\providecommand \@@href[1]{\endgroup#1\@@endlink}%
\providecommand \@sanitize@url [0]{\catcode `\\12\catcode `\$12\catcode
  `\&12\catcode `\#12\catcode `\^12\catcode `\_12\catcode `\%12\relax}%
\providecommand \@@startlink[1]{}%
\providecommand \@@endlink[0]{}%
\providecommand \url  [0]{\begingroup\@sanitize@url \@url }%
\providecommand \@url [1]{\endgroup\@href {#1}{\urlprefix }}%
\providecommand \urlprefix  [0]{URL }%
\providecommand \Eprint [0]{\href }%
\providecommand \doibase [0]{https://doi.org/}%
\providecommand \selectlanguage [0]{\@gobble}%
\providecommand \bibinfo  [0]{\@secondoftwo}%
\providecommand \bibfield  [0]{\@secondoftwo}%
\providecommand \translation [1]{[#1]}%
\providecommand \BibitemOpen [0]{}%
\providecommand \bibitemStop [0]{}%
\providecommand \bibitemNoStop [0]{.\EOS\space}%
\providecommand \EOS [0]{\spacefactor3000\relax}%
\providecommand \BibitemShut  [1]{\csname bibitem#1\endcsname}%
\let\auto@bib@innerbib\@empty
%</preamble>
\bibitem [{\citenamefont {Li}\ \emph {et~al.}(2021)\citenamefont {Li},
  \citenamefont {Zhang}, \citenamefont {Li}, \citenamefont {Fang},\ and\
  \citenamefont {Dong}}]{Li2021}%
  \BibitemOpen
  \bibfield  {author} {\bibinfo {author} {\bibfnamefont {C.}~\bibnamefont
  {Li}}, \bibinfo {author} {\bibfnamefont {X.}~\bibnamefont {Zhang}}, \bibinfo
  {author} {\bibfnamefont {J.}~\bibnamefont {Li}}, \bibinfo {author}
  {\bibfnamefont {T.}~\bibnamefont {Fang}},\ and\ \bibinfo {author}
  {\bibfnamefont {X.}~\bibnamefont {Dong}},\ }\bibfield  {title} {\bibinfo
  {title} {The challenges of modern computing and new opportunities for
  optics},\ }\href {https://doi.org/10.1186/s43074-021-00042-0} {\bibfield
  {journal} {\bibinfo  {journal} {PhotoniX}\ }\textbf {\bibinfo {volume} {2}},\
  \bibinfo {pages} {20} (\bibinfo {year} {2021})}\BibitemShut {NoStop}%
\bibitem [{\citenamefont {Stroev}\ and\ \citenamefont
  {Berloff}(2023)}]{Nikita_AQT2023}%
  \BibitemOpen
  \bibfield  {author} {\bibinfo {author} {\bibfnamefont {N.}~\bibnamefont
  {Stroev}}\ and\ \bibinfo {author} {\bibfnamefont {N.~G.}\ \bibnamefont
  {Berloff}},\ }\bibfield  {title} {\bibinfo {title} {Analog photonics
  computing for information processing, inference, and optimization},\ }\href
  {https://doi.org/https://doi.org/10.1002/qute.202300055} {\bibfield
  {journal} {\bibinfo  {journal} {Advanced Quantum Technologies}\ }\textbf
  {\bibinfo {volume} {6}},\ \bibinfo {pages} {2300055} (\bibinfo {year}
  {2023})}\BibitemShut {NoStop}%
\bibitem [{\citenamefont {Mohseni}\ \emph {et~al.}(2022)\citenamefont
  {Mohseni}, \citenamefont {McMahon},\ and\ \citenamefont
  {Byrnes}}]{Mohseni_REV2022}%
  \BibitemOpen
  \bibfield  {author} {\bibinfo {author} {\bibfnamefont {N.}~\bibnamefont
  {Mohseni}}, \bibinfo {author} {\bibfnamefont {P.~L.}\ \bibnamefont
  {McMahon}},\ and\ \bibinfo {author} {\bibfnamefont {T.}~\bibnamefont
  {Byrnes}},\ }\bibfield  {title} {\bibinfo {title} {Ising machines as hardware
  solvers of combinatorial optimization problems},\ }\href
  {https://doi.org/10.1038/s42254-022-00440-8} {\bibfield  {journal} {\bibinfo
  {journal} {Nature Reviews Physics}\ }\textbf {\bibinfo {volume} {4}},\
  \bibinfo {pages} {363} (\bibinfo {year} {2022})}\BibitemShut {NoStop}%
\bibitem [{\citenamefont {Strogatz}(2015)}]{Strogatz2015}%
  \BibitemOpen
  \bibfield  {author} {\bibinfo {author} {\bibfnamefont {S.~H.}\ \bibnamefont
  {Strogatz}},\ }\href {https://doi.org/10.1201/9780429492563} {\emph {\bibinfo
  {title} {Nonlinear Dynamics and Chaos: With Applications to Physics, Biology,
  Chemistry, and Engineering}}},\ \bibinfo {edition} {2nd}\ ed.\ (\bibinfo
  {publisher} {CRC Press},\ \bibinfo {address} {Boca Raton, FL},\ \bibinfo
  {year} {2015})\BibitemShut {NoStop}%
\bibitem [{\citenamefont {Syed}\ and\ \citenamefont
  {Berloff}(2023)}]{Marvin_2023}%
  \BibitemOpen
  \bibfield  {author} {\bibinfo {author} {\bibfnamefont {M.}~\bibnamefont
  {Syed}}\ and\ \bibinfo {author} {\bibfnamefont {N.~G.}\ \bibnamefont
  {Berloff}},\ }\bibfield  {title} {\bibinfo {title} {Physics-enhanced
  bifurcation optimisers: All you need is a canonical complex network},\ }\href
  {https://doi.org/10.1109/JSTQE.2023.3235334} {\bibfield  {journal} {\bibinfo
  {journal} {IEEE Journal of Selected Topics in Quantum Electronics}\ }\textbf
  {\bibinfo {volume} {29}},\ \bibinfo {pages} {7400406} (\bibinfo {year}
  {2023})}\BibitemShut {NoStop}%
\bibitem [{\citenamefont {Parrondo}\ \emph {et~al.}(2015)\citenamefont
  {Parrondo}, \citenamefont {Horowitz},\ and\ \citenamefont
  {Sagawa}}]{Parrondo_REV2015}%
  \BibitemOpen
  \bibfield  {author} {\bibinfo {author} {\bibfnamefont {J.~M.~R.}\
  \bibnamefont {Parrondo}}, \bibinfo {author} {\bibfnamefont {J.~M.}\
  \bibnamefont {Horowitz}},\ and\ \bibinfo {author} {\bibfnamefont
  {T.}~\bibnamefont {Sagawa}},\ }\bibfield  {title} {\bibinfo {title}
  {Thermodynamics of information},\ }\href {https://doi.org/10.1038/nphys3230}
  {\bibfield  {journal} {\bibinfo  {journal} {Nature Physics}\ }\textbf
  {\bibinfo {volume} {11}},\ \bibinfo {pages} {131} (\bibinfo {year}
  {2015})}\BibitemShut {NoStop}%
\bibitem [{\citenamefont {Vadlamani}\ \emph {et~al.}(2020)\citenamefont
  {Vadlamani}, \citenamefont {Xiao},\ and\ \citenamefont
  {Yablonovitch}}]{Yablonovitch_PNAS2020}%
  \BibitemOpen
  \bibfield  {author} {\bibinfo {author} {\bibfnamefont {S.~K.}\ \bibnamefont
  {Vadlamani}}, \bibinfo {author} {\bibfnamefont {T.~P.}\ \bibnamefont
  {Xiao}},\ and\ \bibinfo {author} {\bibfnamefont {E.}~\bibnamefont
  {Yablonovitch}},\ }\bibfield  {title} {\bibinfo {title} {Physics successfully
  implements lagrange multiplier optimization},\ }\href
  {https://www.jstor.org/stable/26970610} {\bibfield  {journal} {\bibinfo
  {journal} {Proceedings of the National Academy of Sciences of the United
  States of America}\ }\textbf {\bibinfo {volume} {117}},\ \bibinfo {pages}
  {26639} (\bibinfo {year} {2020})}\BibitemShut {NoStop}%
\bibitem [{\citenamefont {Feynman}\ \emph {et~al.}(1964)\citenamefont
  {Feynman}, \citenamefont {Leighton},\ and\ \citenamefont
  {Sands}}]{feynman1964lectures}%
  \BibitemOpen
  \bibfield  {author} {\bibinfo {author} {\bibfnamefont {R.~P.}\ \bibnamefont
  {Feynman}}, \bibinfo {author} {\bibfnamefont {R.~B.}\ \bibnamefont
  {Leighton}},\ and\ \bibinfo {author} {\bibfnamefont {M.}~\bibnamefont
  {Sands}},\ }\href {https://www.feynmanlectures.caltech.edu/} {\emph {\bibinfo
  {title} {The Feynman Lectures on Physics, Volume II}}}\ (\bibinfo
  {publisher} {Addison-Wesley},\ \bibinfo {year} {1964})\BibitemShut {NoStop}%
\bibitem [{\citenamefont {Deng}\ \emph {et~al.}(2010)\citenamefont {Deng},
  \citenamefont {Haug},\ and\ \citenamefont {Yamamoto}}]{Deng_RMP2010}%
  \BibitemOpen
  \bibfield  {author} {\bibinfo {author} {\bibfnamefont {H.}~\bibnamefont
  {Deng}}, \bibinfo {author} {\bibfnamefont {H.}~\bibnamefont {Haug}},\ and\
  \bibinfo {author} {\bibfnamefont {Y.}~\bibnamefont {Yamamoto}},\ }\bibfield
  {title} {\bibinfo {title} {Exciton-polariton {B}ose-{E}instein
  condensation},\ }\href {https://doi.org/10.1103/RevModPhys.82.1489}
  {\bibfield  {journal} {\bibinfo  {journal} {Rev. Mod. Phys.}\ }\textbf
  {\bibinfo {volume} {82}},\ \bibinfo {pages} {1489} (\bibinfo {year}
  {2010})}\BibitemShut {NoStop}%
\bibitem [{\citenamefont {Byrnes}\ \emph {et~al.}(2014)\citenamefont {Byrnes},
  \citenamefont {Kim},\ and\ \citenamefont {Yamamoto}}]{Byrnes2014}%
  \BibitemOpen
  \bibfield  {author} {\bibinfo {author} {\bibfnamefont {T.}~\bibnamefont
  {Byrnes}}, \bibinfo {author} {\bibfnamefont {N.~Y.}\ \bibnamefont {Kim}},\
  and\ \bibinfo {author} {\bibfnamefont {Y.}~\bibnamefont {Yamamoto}},\
  }\bibfield  {title} {\bibinfo {title} {Exciton–polariton condensates},\
  }\href {https://doi.org/10.1038/nphys3143} {\bibfield  {journal} {\bibinfo
  {journal} {Nature Physics}\ }\textbf {\bibinfo {volume} {10}},\ \bibinfo
  {pages} {803} (\bibinfo {year} {2014})}\BibitemShut {NoStop}%
\bibitem [{\citenamefont {Opala}\ and\ \citenamefont
  {Matuszewski}(2023)}]{Opala_REV2023}%
  \BibitemOpen
  \bibfield  {author} {\bibinfo {author} {\bibfnamefont {A.}~\bibnamefont
  {Opala}}\ and\ \bibinfo {author} {\bibfnamefont {M.}~\bibnamefont
  {Matuszewski}},\ }\href@noop {} {\bibinfo {title} {Harnessing
  exciton-polaritons for digital computing, neuromorphic computing, and
  optimization}} (\bibinfo {year} {2023}),\ \Eprint
  {https://arxiv.org/abs/2306.06604} {arXiv:2306.06604} \BibitemShut {NoStop}%
\bibitem [{\citenamefont {Cummins}\ \emph {et~al.}(2025)\citenamefont
  {Cummins}, \citenamefont {Salman},\ and\ \citenamefont
  {Berloff}}]{cummins2023}%
  \BibitemOpen
  \bibfield  {author} {\bibinfo {author} {\bibfnamefont {J.~S.}\ \bibnamefont
  {Cummins}}, \bibinfo {author} {\bibfnamefont {H.}~\bibnamefont {Salman}},\
  and\ \bibinfo {author} {\bibfnamefont {N.~G.}\ \bibnamefont {Berloff}},\
  }\bibfield  {title} {\bibinfo {title} {Ising hamiltonian minimization:
  Gain-based computing with manifold reduction of soft spins vs quantum
  annealing},\ }\href {https://doi.org/10.1103/PhysRevResearch.7.013150}
  {\bibfield  {journal} {\bibinfo  {journal} {Phys. Rev. Res.}\ }\textbf
  {\bibinfo {volume} {7}},\ \bibinfo {pages} {013150} (\bibinfo {year}
  {2025})}\BibitemShut {NoStop}%
\bibitem [{\citenamefont {Opala}\ \emph {et~al.}(2019)\citenamefont {Opala},
  \citenamefont {Ghosh}, \citenamefont {Liew},\ and\ \citenamefont
  {Matuszewski}}]{Opala_PRApplied2019}%
  \BibitemOpen
  \bibfield  {author} {\bibinfo {author} {\bibfnamefont {A.}~\bibnamefont
  {Opala}}, \bibinfo {author} {\bibfnamefont {S.}~\bibnamefont {Ghosh}},
  \bibinfo {author} {\bibfnamefont {T.~C.}\ \bibnamefont {Liew}},\ and\
  \bibinfo {author} {\bibfnamefont {M.}~\bibnamefont {Matuszewski}},\
  }\bibfield  {title} {\bibinfo {title} {Neuromorphic computing in
  {G}inzburg-{L}andau polariton-lattice systems},\ }\href
  {https://doi.org/10.1103/PhysRevApplied.11.064029} {\bibfield  {journal}
  {\bibinfo  {journal} {Phys. Rev. Appl.}\ }\textbf {\bibinfo {volume} {11}},\
  \bibinfo {pages} {064029} (\bibinfo {year} {2019})}\BibitemShut {NoStop}%
\bibitem [{\citenamefont {Sedov}\ and\ \citenamefont
  {Kavokin}(2025)}]{Sedov_2025}%
  \BibitemOpen
  \bibfield  {author} {\bibinfo {author} {\bibfnamefont {E.}~\bibnamefont
  {Sedov}}\ and\ \bibinfo {author} {\bibfnamefont {A.}~\bibnamefont
  {Kavokin}},\ }\bibfield  {title} {\bibinfo {title} {Polariton lattices as
  binarized neuromorphic networks},\ }\href
  {http://dx.doi.org/10.1038/s41377-024-01719-4} {\bibfield  {journal}
  {\bibinfo  {journal} {Light: Science $\&$ Applications}\ }\textbf {\bibinfo
  {volume} {14}},\ \bibinfo {pages} {52} (\bibinfo {year} {2025})}\BibitemShut
  {NoStop}%
\bibitem [{\citenamefont {McMahon}\ \emph {et~al.}(2016)\citenamefont
  {McMahon}, \citenamefont {Marandi}, \citenamefont {Haribara}, \citenamefont
  {Hamerly}, \citenamefont {Langrock}, \citenamefont {Tamate}, \citenamefont
  {Inagaki}, \citenamefont {Takesue}, \citenamefont {Utsunomiya}, \citenamefont
  {Aihara}, \citenamefont {Byer}, \citenamefont {Fejer}, \citenamefont
  {Mabuchi},\ and\ \citenamefont {Yamamoto}}]{Peter_Science2016}%
  \BibitemOpen
  \bibfield  {author} {\bibinfo {author} {\bibfnamefont {P.~L.}\ \bibnamefont
  {McMahon}}, \bibinfo {author} {\bibfnamefont {A.}~\bibnamefont {Marandi}},
  \bibinfo {author} {\bibfnamefont {Y.}~\bibnamefont {Haribara}}, \bibinfo
  {author} {\bibfnamefont {R.}~\bibnamefont {Hamerly}}, \bibinfo {author}
  {\bibfnamefont {C.}~\bibnamefont {Langrock}}, \bibinfo {author}
  {\bibfnamefont {S.}~\bibnamefont {Tamate}}, \bibinfo {author} {\bibfnamefont
  {T.}~\bibnamefont {Inagaki}}, \bibinfo {author} {\bibfnamefont
  {H.}~\bibnamefont {Takesue}}, \bibinfo {author} {\bibfnamefont
  {S.}~\bibnamefont {Utsunomiya}}, \bibinfo {author} {\bibfnamefont
  {K.}~\bibnamefont {Aihara}}, \bibinfo {author} {\bibfnamefont {R.~L.}\
  \bibnamefont {Byer}}, \bibinfo {author} {\bibfnamefont {M.~M.}\ \bibnamefont
  {Fejer}}, \bibinfo {author} {\bibfnamefont {H.}~\bibnamefont {Mabuchi}},\
  and\ \bibinfo {author} {\bibfnamefont {Y.}~\bibnamefont {Yamamoto}},\
  }\bibfield  {title} {\bibinfo {title} {A fully programmable 100-spin coherent
  {I}sing machine with all-to-all connections},\ }\href
  {https://doi.org/10.1126/science.aah5178} {\bibfield  {journal} {\bibinfo
  {journal} {Science}\ }\textbf {\bibinfo {volume} {354}},\ \bibinfo {pages}
  {614} (\bibinfo {year} {2016})}\BibitemShut {NoStop}%
\bibitem [{\citenamefont {Yamamoto}\ \emph {et~al.}(2020)\citenamefont
  {Yamamoto}, \citenamefont {Leleu}, \citenamefont {Ganguli},\ and\
  \citenamefont {Mabuchi}}]{Yamamoto_REV2020}%
  \BibitemOpen
  \bibfield  {author} {\bibinfo {author} {\bibfnamefont {Y.}~\bibnamefont
  {Yamamoto}}, \bibinfo {author} {\bibfnamefont {T.}~\bibnamefont {Leleu}},
  \bibinfo {author} {\bibfnamefont {S.}~\bibnamefont {Ganguli}},\ and\ \bibinfo
  {author} {\bibfnamefont {H.}~\bibnamefont {Mabuchi}},\ }\bibfield  {title}
  {\bibinfo {title} {Coherent {I}sing machines—quantum optics and neural
  network perspectives},\ }\href {https://doi.org/10.1063/5.0016140} {\bibfield
   {journal} {\bibinfo  {journal} {Applied Physics Letters}\ }\textbf {\bibinfo
  {volume} {117}},\ \bibinfo {pages} {160501} (\bibinfo {year}
  {2020})}\BibitemShut {NoStop}%
\bibitem [{\citenamefont {Gao}\ \emph {et~al.}(2024)\citenamefont {Gao},
  \citenamefont {Chen}, \citenamefont {Qi}, \citenamefont {Fu}, \citenamefont
  {Yuan},\ and\ \citenamefont {Danner}}]{Gao_REV2024}%
  \BibitemOpen
  \bibfield  {author} {\bibinfo {author} {\bibfnamefont {Y.}~\bibnamefont
  {Gao}}, \bibinfo {author} {\bibfnamefont {G.}~\bibnamefont {Chen}}, \bibinfo
  {author} {\bibfnamefont {L.}~\bibnamefont {Qi}}, \bibinfo {author}
  {\bibfnamefont {W.}~\bibnamefont {Fu}}, \bibinfo {author} {\bibfnamefont
  {Z.}~\bibnamefont {Yuan}},\ and\ \bibinfo {author} {\bibfnamefont {A.~J.}\
  \bibnamefont {Danner}},\ }\bibfield  {title} {\bibinfo {title} {Photonic
  {I}sing machines for combinatorial optimization problems},\ }\href
  {https://doi.org/10.1063/5.0216656} {\bibfield  {journal} {\bibinfo
  {journal} {Applied Physics Reviews}\ }\textbf {\bibinfo {volume} {11}},\
  \bibinfo {pages} {041307} (\bibinfo {year} {2024})}\BibitemShut {NoStop}%
\bibitem [{\citenamefont {Shastri}\ \emph {et~al.}(2021)\citenamefont
  {Shastri}, \citenamefont {Tait}, \citenamefont {de~Lima}, \citenamefont
  {Pernice}, \citenamefont {Bhaskaran}, \citenamefont {Wright},\ and\
  \citenamefont {Prucnal}}]{Shastri2021}%
  \BibitemOpen
  \bibfield  {author} {\bibinfo {author} {\bibfnamefont {B.~J.}\ \bibnamefont
  {Shastri}}, \bibinfo {author} {\bibfnamefont {A.~N.}\ \bibnamefont {Tait}},
  \bibinfo {author} {\bibfnamefont {T.~F.}\ \bibnamefont {de~Lima}}, \bibinfo
  {author} {\bibfnamefont {W.~H.~P.}\ \bibnamefont {Pernice}}, \bibinfo
  {author} {\bibfnamefont {H.}~\bibnamefont {Bhaskaran}}, \bibinfo {author}
  {\bibfnamefont {C.~D.}\ \bibnamefont {Wright}},\ and\ \bibinfo {author}
  {\bibfnamefont {P.~R.}\ \bibnamefont {Prucnal}},\ }\bibfield  {title}
  {\bibinfo {title} {Photonics for artificial intelligence and neuromorphic
  computing},\ }\href {https://doi.org/10.1038/s41566-020-00754-y} {\bibfield
  {journal} {\bibinfo  {journal} {Nature Photonics}\ }\textbf {\bibinfo
  {volume} {15}},\ \bibinfo {pages} {102} (\bibinfo {year} {2021})}\BibitemShut
  {NoStop}%
\bibitem [{\citenamefont {Wu}\ \emph {et~al.}(2022)\citenamefont {Wu},
  \citenamefont {Lin}, \citenamefont {Guo}, \citenamefont {Liu}, \citenamefont
  {Fang}, \citenamefont {Jiao},\ and\ \citenamefont {Dai}}]{WU2022133}%
  \BibitemOpen
  \bibfield  {author} {\bibinfo {author} {\bibfnamefont {J.}~\bibnamefont
  {Wu}}, \bibinfo {author} {\bibfnamefont {X.}~\bibnamefont {Lin}}, \bibinfo
  {author} {\bibfnamefont {Y.}~\bibnamefont {Guo}}, \bibinfo {author}
  {\bibfnamefont {J.}~\bibnamefont {Liu}}, \bibinfo {author} {\bibfnamefont
  {L.}~\bibnamefont {Fang}}, \bibinfo {author} {\bibfnamefont {S.}~\bibnamefont
  {Jiao}},\ and\ \bibinfo {author} {\bibfnamefont {Q.}~\bibnamefont {Dai}},\
  }\bibfield  {title} {\bibinfo {title} {Analog optical computing for
  artificial intelligence},\ }\href
  {https://doi.org/https://doi.org/10.1016/j.eng.2021.06.021} {\bibfield
  {journal} {\bibinfo  {journal} {Engineering}\ }\textbf {\bibinfo {volume}
  {10}},\ \bibinfo {pages} {133} (\bibinfo {year} {2022})}\BibitemShut
  {NoStop}%
\bibitem [{\citenamefont {Prabhu}\ \emph {et~al.}(2020)\citenamefont {Prabhu},
  \citenamefont {Roques-Carmes}, \citenamefont {Shen}, \citenamefont {Harris},
  \citenamefont {Jing}, \citenamefont {Carolan}, \citenamefont {Hamerly},
  \citenamefont {Baehr-Jones}, \citenamefont {Hochberg}, \citenamefont
  {\v{C}eperi\'{c}}, \citenamefont {Joannopoulos}, \citenamefont {Englund},\
  and\ \citenamefont {Solja\v{c}i\'{c}}}]{Prabhu_Optica2020}%
  \BibitemOpen
  \bibfield  {author} {\bibinfo {author} {\bibfnamefont {M.}~\bibnamefont
  {Prabhu}}, \bibinfo {author} {\bibfnamefont {C.}~\bibnamefont
  {Roques-Carmes}}, \bibinfo {author} {\bibfnamefont {Y.}~\bibnamefont {Shen}},
  \bibinfo {author} {\bibfnamefont {N.}~\bibnamefont {Harris}}, \bibinfo
  {author} {\bibfnamefont {L.}~\bibnamefont {Jing}}, \bibinfo {author}
  {\bibfnamefont {J.}~\bibnamefont {Carolan}}, \bibinfo {author} {\bibfnamefont
  {R.}~\bibnamefont {Hamerly}}, \bibinfo {author} {\bibfnamefont
  {T.}~\bibnamefont {Baehr-Jones}}, \bibinfo {author} {\bibfnamefont
  {M.}~\bibnamefont {Hochberg}}, \bibinfo {author} {\bibfnamefont
  {V.}~\bibnamefont {\v{C}eperi\'{c}}}, \bibinfo {author} {\bibfnamefont
  {J.~D.}\ \bibnamefont {Joannopoulos}}, \bibinfo {author} {\bibfnamefont
  {D.~R.}\ \bibnamefont {Englund}},\ and\ \bibinfo {author} {\bibfnamefont
  {M.}~\bibnamefont {Solja\v{c}i\'{c}}},\ }\bibfield  {title} {\bibinfo {title}
  {Accelerating recurrent {I}sing machines in photonic integrated circuits},\
  }\href {https://doi.org/10.1364/OPTICA.386613} {\bibfield  {journal}
  {\bibinfo  {journal} {Optica}\ }\textbf {\bibinfo {volume} {7}},\ \bibinfo
  {pages} {551} (\bibinfo {year} {2020})}\BibitemShut {NoStop}%
\bibitem [{\citenamefont {Berloff}\ \emph {et~al.}(2017)\citenamefont
  {Berloff}, \citenamefont {Silva}, \citenamefont {Kalinin}, \citenamefont
  {Askitopoulos}, \citenamefont {Töpfer}, \citenamefont {Cilibrizzi},
  \citenamefont {Langbein},\ and\ \citenamefont {Lagoudakis}}]{Berloff2017}%
  \BibitemOpen
  \bibfield  {author} {\bibinfo {author} {\bibfnamefont {N.~G.}\ \bibnamefont
  {Berloff}}, \bibinfo {author} {\bibfnamefont {M.}~\bibnamefont {Silva}},
  \bibinfo {author} {\bibfnamefont {K.}~\bibnamefont {Kalinin}}, \bibinfo
  {author} {\bibfnamefont {A.}~\bibnamefont {Askitopoulos}}, \bibinfo {author}
  {\bibfnamefont {J.~D.}\ \bibnamefont {Töpfer}}, \bibinfo {author}
  {\bibfnamefont {P.}~\bibnamefont {Cilibrizzi}}, \bibinfo {author}
  {\bibfnamefont {W.}~\bibnamefont {Langbein}},\ and\ \bibinfo {author}
  {\bibfnamefont {P.~G.}\ \bibnamefont {Lagoudakis}},\ }\bibfield  {title}
  {\bibinfo {title} {Realizing the classical {XY} {Hamiltonian} in polariton
  simulators},\ }\href {https://doi.org/10.1038/nmat4971} {\bibfield  {journal}
  {\bibinfo  {journal} {Nature Materials}\ }\textbf {\bibinfo {volume} {16}},\
  \bibinfo {pages} {1120} (\bibinfo {year} {2017})}\BibitemShut {NoStop}%
\bibitem [{\citenamefont {Peng}\ \emph {et~al.}(2024)\citenamefont {Peng},
  \citenamefont {Li}, \citenamefont {Berloff}, \citenamefont {Zhang},\ and\
  \citenamefont {Bao}}]{Peng_NanoPho2024}%
  \BibitemOpen
  \bibfield  {author} {\bibinfo {author} {\bibfnamefont {K.}~\bibnamefont
  {Peng}}, \bibinfo {author} {\bibfnamefont {W.}~\bibnamefont {Li}}, \bibinfo
  {author} {\bibfnamefont {N.~G.}\ \bibnamefont {Berloff}}, \bibinfo {author}
  {\bibfnamefont {X.}~\bibnamefont {Zhang}},\ and\ \bibinfo {author}
  {\bibfnamefont {W.}~\bibnamefont {Bao}},\ }\bibfield  {title} {\bibinfo
  {title} {Room temperature polaritonic soft-spin {XY} {Hamiltonian} in
  organic–inorganic halide perovskites},\ }\href
  {https://doi.org/doi:10.1515/nanoph-2023-0818} {\bibfield  {journal}
  {\bibinfo  {journal} {Nanophotonics}\ }\textbf {\bibinfo {volume} {13}},\
  \bibinfo {pages} {2651} (\bibinfo {year} {2024})}\BibitemShut {NoStop}%
\bibitem [{\citenamefont {Zhang}(2020)}]{Zhang_PRE2020}%
  \BibitemOpen
  \bibfield  {author} {\bibinfo {author} {\bibfnamefont {B.}~\bibnamefont
  {Zhang}},\ }\bibfield  {title} {\bibinfo {title} {Perturbative study of the
  one-dimensional quantum clock model},\ }\href
  {https://doi.org/10.1103/PhysRevE.102.042110} {\bibfield  {journal} {\bibinfo
   {journal} {Phys. Rev. E}\ }\textbf {\bibinfo {volume} {102}},\ \bibinfo
  {pages} {042110} (\bibinfo {year} {2020})}\BibitemShut {NoStop}%
\bibitem [{\citenamefont {Bairey}\ \emph {et~al.}(2019)\citenamefont {Bairey},
  \citenamefont {Arad},\ and\ \citenamefont {Lindner}}]{Bairey_PRL2019}%
  \BibitemOpen
  \bibfield  {author} {\bibinfo {author} {\bibfnamefont {E.}~\bibnamefont
  {Bairey}}, \bibinfo {author} {\bibfnamefont {I.}~\bibnamefont {Arad}},\ and\
  \bibinfo {author} {\bibfnamefont {N.~H.}\ \bibnamefont {Lindner}},\
  }\bibfield  {title} {\bibinfo {title} {Learning a local {Hamiltonian} from
  local measurements},\ }\href {https://doi.org/10.1103/PhysRevLett.122.020504}
  {\bibfield  {journal} {\bibinfo  {journal} {Phys. Rev. Lett.}\ }\textbf
  {\bibinfo {volume} {122}},\ \bibinfo {pages} {020504} (\bibinfo {year}
  {2019})}\BibitemShut {NoStop}%
\bibitem [{\citenamefont {Kalinin}\ \emph {et~al.}(2023)\citenamefont
  {Kalinin}, \citenamefont {Mourgias-Alexandris}, \citenamefont {Ballani},
  \citenamefont {Berloff}, \citenamefont {Clegg}, \citenamefont {Cletheroe},
  \citenamefont {Gkantsidis}, \citenamefont {Haller}, \citenamefont
  {Lyutsarev}, \citenamefont {Parmigiani}, \citenamefont {Pickup},\ and\
  \citenamefont {Rowstron}}]{kalinin2023}%
  \BibitemOpen
  \bibfield  {author} {\bibinfo {author} {\bibfnamefont {K.~P.}\ \bibnamefont
  {Kalinin}}, \bibinfo {author} {\bibfnamefont {G.}~\bibnamefont
  {Mourgias-Alexandris}}, \bibinfo {author} {\bibfnamefont {H.}~\bibnamefont
  {Ballani}}, \bibinfo {author} {\bibfnamefont {N.~G.}\ \bibnamefont
  {Berloff}}, \bibinfo {author} {\bibfnamefont {J.~H.}\ \bibnamefont {Clegg}},
  \bibinfo {author} {\bibfnamefont {D.}~\bibnamefont {Cletheroe}}, \bibinfo
  {author} {\bibfnamefont {C.}~\bibnamefont {Gkantsidis}}, \bibinfo {author}
  {\bibfnamefont {I.}~\bibnamefont {Haller}}, \bibinfo {author} {\bibfnamefont
  {V.}~\bibnamefont {Lyutsarev}}, \bibinfo {author} {\bibfnamefont
  {F.}~\bibnamefont {Parmigiani}}, \bibinfo {author} {\bibfnamefont
  {L.}~\bibnamefont {Pickup}},\ and\ \bibinfo {author} {\bibfnamefont
  {A.}~\bibnamefont {Rowstron}},\ }\href@noop {} {\bibinfo {title} {Analog
  iterative machine ({AIM}): {Using} light to solve quadratic optimization
  problems with mixed variables}} (\bibinfo {year} {2023}),\ \Eprint
  {https://arxiv.org/abs/2304.12594} {arXiv:2304.12594} \BibitemShut {NoStop}%
\bibitem [{\citenamefont {Khosravi}\ \emph {et~al.}(2023)\citenamefont
  {Khosravi}, \citenamefont {Yildiz}, \citenamefont {Scherer},\ and\
  \citenamefont {Ronagh}}]{khosravi2023}%
  \BibitemOpen
  \bibfield  {author} {\bibinfo {author} {\bibfnamefont {F.}~\bibnamefont
  {Khosravi}}, \bibinfo {author} {\bibfnamefont {U.}~\bibnamefont {Yildiz}},
  \bibinfo {author} {\bibfnamefont {A.}~\bibnamefont {Scherer}},\ and\ \bibinfo
  {author} {\bibfnamefont {P.}~\bibnamefont {Ronagh}},\ }\href@noop {}
  {\bibinfo {title} {Non-convex quadratic programming using coherent optical
  networks}} (\bibinfo {year} {2023}),\ \Eprint
  {https://arxiv.org/abs/2209.04415} {arXiv:2209.04415} \BibitemShut {NoStop}%
\bibitem [{\citenamefont {Kamaletdinov}\ and\ \citenamefont
  {Berloff}(2024)}]{kamaletdinov2024coupling}%
  \BibitemOpen
  \bibfield  {author} {\bibinfo {author} {\bibfnamefont {A.}~\bibnamefont
  {Kamaletdinov}}\ and\ \bibinfo {author} {\bibfnamefont {N.~G.}\ \bibnamefont
  {Berloff}},\ }\href@noop {} {\bibinfo {title} {Coupling light with matter for
  identifying dominant subnetworks}} (\bibinfo {year} {2024}),\ \Eprint
  {https://arxiv.org/abs/2405.17296} {arXiv:2405.17296} \BibitemShut {NoStop}%
\bibitem [{\citenamefont {Wang}\ \emph {et~al.}(2025)\citenamefont {Wang},
  \citenamefont {Cummins}, \citenamefont {Syed}, \citenamefont {Stroev},
  \citenamefont {Pastras}, \citenamefont {Sakellariou}, \citenamefont
  {Tsintzos}, \citenamefont {Askitopoulos}, \citenamefont {Veraldi},
  \citenamefont {Calvanese~Strinati}, \citenamefont {Gentilini}, \citenamefont
  {Pierangeli}, \citenamefont {Conti},\ and\ \citenamefont
  {Berloff}}]{Wang2025}%
  \BibitemOpen
  \bibfield  {author} {\bibinfo {author} {\bibfnamefont {R.~Z.}\ \bibnamefont
  {Wang}}, \bibinfo {author} {\bibfnamefont {J.~S.}\ \bibnamefont {Cummins}},
  \bibinfo {author} {\bibfnamefont {M.}~\bibnamefont {Syed}}, \bibinfo {author}
  {\bibfnamefont {N.}~\bibnamefont {Stroev}}, \bibinfo {author} {\bibfnamefont
  {G.}~\bibnamefont {Pastras}}, \bibinfo {author} {\bibfnamefont
  {J.}~\bibnamefont {Sakellariou}}, \bibinfo {author} {\bibfnamefont
  {S.}~\bibnamefont {Tsintzos}}, \bibinfo {author} {\bibfnamefont
  {A.}~\bibnamefont {Askitopoulos}}, \bibinfo {author} {\bibfnamefont
  {D.}~\bibnamefont {Veraldi}}, \bibinfo {author} {\bibfnamefont
  {M.}~\bibnamefont {Calvanese~Strinati}}, \bibinfo {author} {\bibfnamefont
  {S.}~\bibnamefont {Gentilini}}, \bibinfo {author} {\bibfnamefont
  {D.}~\bibnamefont {Pierangeli}}, \bibinfo {author} {\bibfnamefont
  {C.}~\bibnamefont {Conti}},\ and\ \bibinfo {author} {\bibfnamefont {N.~G.}\
  \bibnamefont {Berloff}},\ }\bibfield  {title} {\bibinfo {title} {Efficient
  computation using spatial-photonic {Ising} machines with low-rank and
  circulant matrix constraints},\ }\href
  {https://doi.org/10.1038/s42005-025-01987-5} {\bibfield  {journal} {\bibinfo
  {journal} {Communications Physics}\ }\textbf {\bibinfo {volume} {8}},\
  \bibinfo {pages} {86} (\bibinfo {year} {2025})}\BibitemShut {NoStop}%
\bibitem [{\citenamefont {Dokmani{\'c}}\ \emph {et~al.}(2015)\citenamefont
  {Dokmani{\'c}}, \citenamefont {Parhizkar}, \citenamefont {Ranieri},\ and\
  \citenamefont {Vetterli}}]{Dokmani_REV2015}%
  \BibitemOpen
  \bibfield  {author} {\bibinfo {author} {\bibfnamefont {I.}~\bibnamefont
  {Dokmani{\'c}}}, \bibinfo {author} {\bibfnamefont {R.}~\bibnamefont
  {Parhizkar}}, \bibinfo {author} {\bibfnamefont {J.}~\bibnamefont {Ranieri}},\
  and\ \bibinfo {author} {\bibfnamefont {M.}~\bibnamefont {Vetterli}},\
  }\bibfield  {title} {\bibinfo {title} {Euclidean distance matrices: Essential
  theory, algorithms, and applications},\ }\href
  {https://api.semanticscholar.org/CorpusID:8603398} {\bibfield  {journal}
  {\bibinfo  {journal} {IEEE Signal Processing Magazine}\ }\textbf {\bibinfo
  {volume} {32}},\ \bibinfo {pages} {12} (\bibinfo {year} {2015})}\BibitemShut
  {NoStop}%
\bibitem [{\citenamefont {Liberti}\ \emph {et~al.}(2014)\citenamefont
  {Liberti}, \citenamefont {Lavor}, \citenamefont {Maculan},\ and\
  \citenamefont {Mucherino}}]{Liberti_DistanceREV}%
  \BibitemOpen
  \bibfield  {author} {\bibinfo {author} {\bibfnamefont {L.}~\bibnamefont
  {Liberti}}, \bibinfo {author} {\bibfnamefont {C.}~\bibnamefont {Lavor}},
  \bibinfo {author} {\bibfnamefont {N.}~\bibnamefont {Maculan}},\ and\ \bibinfo
  {author} {\bibfnamefont {A.}~\bibnamefont {Mucherino}},\ }\bibfield  {title}
  {\bibinfo {title} {Euclidean distance geometry and applications},\ }\href
  {https://doi.org/10.1137/120875909} {\bibfield  {journal} {\bibinfo
  {journal} {SIAM Review}\ }\textbf {\bibinfo {volume} {56}},\ \bibinfo {pages}
  {3} (\bibinfo {year} {2014})}\BibitemShut {NoStop}%
\bibitem [{\citenamefont {Aspnes}\ \emph {et~al.}(2006)\citenamefont {Aspnes},
  \citenamefont {Eren}, \citenamefont {Goldenberg}, \citenamefont {Morse},
  \citenamefont {Whiteley}, \citenamefont {Yang}, \citenamefont {Anderson},\
  and\ \citenamefont {Belhumeur}}]{AspnesIEEE2006}%
  \BibitemOpen
  \bibfield  {author} {\bibinfo {author} {\bibfnamefont {J.}~\bibnamefont
  {Aspnes}}, \bibinfo {author} {\bibfnamefont {T.}~\bibnamefont {Eren}},
  \bibinfo {author} {\bibfnamefont {D.}~\bibnamefont {Goldenberg}}, \bibinfo
  {author} {\bibfnamefont {A.}~\bibnamefont {Morse}}, \bibinfo {author}
  {\bibfnamefont {W.}~\bibnamefont {Whiteley}}, \bibinfo {author}
  {\bibfnamefont {Y.}~\bibnamefont {Yang}}, \bibinfo {author} {\bibfnamefont
  {B.}~\bibnamefont {Anderson}},\ and\ \bibinfo {author} {\bibfnamefont
  {P.}~\bibnamefont {Belhumeur}},\ }\bibfield  {title} {\bibinfo {title} {A
  theory of network localization},\ }\href
  {https://doi.org/10.1109/TMC.2006.174} {\bibfield  {journal} {\bibinfo
  {journal} {IEEE Transactions on Mobile Computing}\ }\textbf {\bibinfo
  {volume} {5}},\ \bibinfo {pages} {1663} (\bibinfo {year} {2006})}\BibitemShut
  {NoStop}%
\bibitem [{\citenamefont {Aspnes}\ \emph {et~al.}(2004)\citenamefont {Aspnes},
  \citenamefont {Goldenberg},\ and\ \citenamefont {Yang}}]{AspnesSNLNP}%
  \BibitemOpen
  \bibfield  {author} {\bibinfo {author} {\bibfnamefont {J.}~\bibnamefont
  {Aspnes}}, \bibinfo {author} {\bibfnamefont {D.}~\bibnamefont {Goldenberg}},\
  and\ \bibinfo {author} {\bibfnamefont {Y.~R.}\ \bibnamefont {Yang}},\
  }\bibfield  {title} {\bibinfo {title} {On the computational complexity of
  sensor network localization},\ }in\ \href
  {https://link.springer.com/chapter/10.1007/978-3-540-27820-7_5} {\emph
  {\bibinfo {booktitle} {Algorithmic Aspects of Wireless Sensor Networks}}},\
  \bibinfo {editor} {edited by\ \bibinfo {editor} {\bibfnamefont {S.~E.}\
  \bibnamefont {Nikoletseas}}\ and\ \bibinfo {editor} {\bibfnamefont
  {J.~D.~P.}\ \bibnamefont {Rolim}}}\ (\bibinfo  {publisher} {Springer Berlin
  Heidelberg},\ \bibinfo {address} {Berlin, Heidelberg},\ \bibinfo {year}
  {2004})\ pp.\ \bibinfo {pages} {32--44}\BibitemShut {NoStop}%
\bibitem [{\citenamefont {Tabassum}\ \emph {et~al.}(2018)\citenamefont
  {Tabassum}, \citenamefont {Pereira}, \citenamefont {Fernandes},\ and\
  \citenamefont {Gama}}]{Social_network_analysis}%
  \BibitemOpen
  \bibfield  {author} {\bibinfo {author} {\bibfnamefont {S.}~\bibnamefont
  {Tabassum}}, \bibinfo {author} {\bibfnamefont {F.~S.~F.}\ \bibnamefont
  {Pereira}}, \bibinfo {author} {\bibfnamefont {S.}~\bibnamefont {Fernandes}},\
  and\ \bibinfo {author} {\bibfnamefont {J.}~\bibnamefont {Gama}},\ }\bibfield
  {title} {\bibinfo {title} {Social network analysis: An overview},\ }\href
  {https://doi.org/https://doi.org/10.1002/widm.1256} {\bibfield  {journal}
  {\bibinfo  {journal} {WIREs Data Mining and Knowledge Discovery}\ }\textbf
  {\bibinfo {volume} {8}},\ \bibinfo {pages} {e1256} (\bibinfo {year}
  {2018})}\BibitemShut {NoStop}%
\bibitem [{\citenamefont {Foedermayr}\ and\ \citenamefont
  {and}(2008)}]{Foedermayr01072008}%
  \BibitemOpen
  \bibfield  {author} {\bibinfo {author} {\bibfnamefont {E.~K.}\ \bibnamefont
  {Foedermayr}}\ and\ \bibinfo {author} {\bibfnamefont {A.~D.}\ \bibnamefont
  {and}},\ }\bibfield  {title} {\bibinfo {title} {Market segmentation in
  practice: Review of empirical studies, methodological assessment, and agenda
  for future research},\ }\href {https://doi.org/10.1080/09652540802117140}
  {\bibfield  {journal} {\bibinfo  {journal} {Journal of Strategic Marketing}\
  }\textbf {\bibinfo {volume} {16}},\ \bibinfo {pages} {223} (\bibinfo {year}
  {2008})}\BibitemShut {NoStop}%
\bibitem [{\citenamefont {Huang}\ \emph {et~al.}(2023)\citenamefont {Huang},
  \citenamefont {Kong}, \citenamefont {Wang}, \citenamefont {Ju}, \citenamefont
  {Zhang}, \citenamefont {Zhu}, \citenamefont {Gong}, \citenamefont {Zhang},
  \citenamefont {Yu}, \citenamefont {Zheng},\ and\ \citenamefont
  {Bu}}]{Protein_Structure}%
  \BibitemOpen
  \bibfield  {author} {\bibinfo {author} {\bibfnamefont {B.}~\bibnamefont
  {Huang}}, \bibinfo {author} {\bibfnamefont {L.}~\bibnamefont {Kong}},
  \bibinfo {author} {\bibfnamefont {C.}~\bibnamefont {Wang}}, \bibinfo {author}
  {\bibfnamefont {F.}~\bibnamefont {Ju}}, \bibinfo {author} {\bibfnamefont
  {Q.}~\bibnamefont {Zhang}}, \bibinfo {author} {\bibfnamefont
  {J.}~\bibnamefont {Zhu}}, \bibinfo {author} {\bibfnamefont {T.}~\bibnamefont
  {Gong}}, \bibinfo {author} {\bibfnamefont {H.}~\bibnamefont {Zhang}},
  \bibinfo {author} {\bibfnamefont {C.}~\bibnamefont {Yu}}, \bibinfo {author}
  {\bibfnamefont {W.-M.}\ \bibnamefont {Zheng}},\ and\ \bibinfo {author}
  {\bibfnamefont {D.}~\bibnamefont {Bu}},\ }\bibfield  {title} {\bibinfo
  {title} {Protein structure prediction: Challenges, advances, and the shift of
  research paradigms},\ }\href {https://doi.org/10.1016/j.gpb.2022.11.014}
  {\bibfield  {journal} {\bibinfo  {journal} {Genomics, Proteomics and
  Bioinformatics}\ }\textbf {\bibinfo {volume} {21}},\ \bibinfo {pages} {913}
  (\bibinfo {year} {2023})}\BibitemShut {NoStop}%
\bibitem [{\citenamefont {Elton}\ \emph {et~al.}(2019)\citenamefont {Elton},
  \citenamefont {Boukouvalas}, \citenamefont {Fuge},\ and\ \citenamefont
  {Chung}}]{molecular_design}%
  \BibitemOpen
  \bibfield  {author} {\bibinfo {author} {\bibfnamefont {D.~C.}\ \bibnamefont
  {Elton}}, \bibinfo {author} {\bibfnamefont {Z.}~\bibnamefont {Boukouvalas}},
  \bibinfo {author} {\bibfnamefont {M.~D.}\ \bibnamefont {Fuge}},\ and\
  \bibinfo {author} {\bibfnamefont {P.~W.}\ \bibnamefont {Chung}},\ }\bibfield
  {title} {\bibinfo {title} {Deep learning for molecular design—a review of
  the state of the art},\ }\href {https://doi.org/10.1039/C9ME00039A}
  {\bibfield  {journal} {\bibinfo  {journal} {Molecular Systems Design and
  Engineering}\ }\textbf {\bibinfo {volume} {4}},\ \bibinfo {pages} {828}
  (\bibinfo {year} {2019})}\BibitemShut {NoStop}%
\bibitem [{\citenamefont {Wuestefeld}\ \emph {et~al.}(2018)\citenamefont
  {Wuestefeld}, \citenamefont {Greve}, \citenamefont {Näsholm},\ and\
  \citenamefont {Oye}}]{earthquake_location}%
  \BibitemOpen
  \bibfield  {author} {\bibinfo {author} {\bibfnamefont {A.}~\bibnamefont
  {Wuestefeld}}, \bibinfo {author} {\bibfnamefont {S.~M.}\ \bibnamefont
  {Greve}}, \bibinfo {author} {\bibfnamefont {S.~P.}\ \bibnamefont
  {Näsholm}},\ and\ \bibinfo {author} {\bibfnamefont {V.}~\bibnamefont
  {Oye}},\ }\bibfield  {title} {\bibinfo {title} {Benchmarking earthquake
  location algorithms: A synthetic comparison},\ }\href
  {https://doi.org/10.1190/geo2017-0317.1} {\bibfield  {journal} {\bibinfo
  {journal} {GEOPHYSICS}\ }\textbf {\bibinfo {volume} {83}},\ \bibinfo {pages}
  {KS35} (\bibinfo {year} {2018})}\BibitemShut {NoStop}%
\bibitem [{\citenamefont {Strang}(2019)}]{strang2019}%
  \BibitemOpen
  \bibfield  {author} {\bibinfo {author} {\bibfnamefont {G.}~\bibnamefont
  {Strang}},\ }\href
  {https://www.cambridge.org/gb/universitypress/subjects/computer-science/pattern-recognition-and-machine-learning/linear-algebra-and-learning-data}
  {\emph {\bibinfo {title} {Linear Algebra and Learning from Data}}}\ (\bibinfo
   {publisher} {Cambridge University Press},\ \bibinfo {year}
  {2019})\BibitemShut {NoStop}%
\bibitem [{\citenamefont {Parhizkar}(2013)}]{Parhizkar_thesis}%
  \BibitemOpen
  \bibfield  {author} {\bibinfo {author} {\bibfnamefont {R.}~\bibnamefont
  {Parhizkar}},\ }\emph {\bibinfo {title} {Euclidean Distance Matrices:
  Properties, Algorithms and Applications}},\ \href
  {https://doi.org/10.5075/epfl-thesis-5971} {Ph.D. thesis},\ \bibinfo
  {school} {EPFL}, \bibinfo {address} {Lausanne} (\bibinfo {year}
  {2013})\BibitemShut {NoStop}%
\bibitem [{\citenamefont {Biswas}\ \emph {et~al.}(2006)\citenamefont {Biswas},
  \citenamefont {Liang}, \citenamefont {Toh}, \citenamefont {Ye},\ and\
  \citenamefont {Wang}}]{BiswasIEEE2006}%
  \BibitemOpen
  \bibfield  {author} {\bibinfo {author} {\bibfnamefont {P.}~\bibnamefont
  {Biswas}}, \bibinfo {author} {\bibfnamefont {T.-C.}\ \bibnamefont {Liang}},
  \bibinfo {author} {\bibfnamefont {K.-C.}\ \bibnamefont {Toh}}, \bibinfo
  {author} {\bibfnamefont {Y.}~\bibnamefont {Ye}},\ and\ \bibinfo {author}
  {\bibfnamefont {T.-C.}\ \bibnamefont {Wang}},\ }\bibfield  {title} {\bibinfo
  {title} {Semidefinite programming approaches for sensor network localization
  with noisy distance measurements},\ }\href
  {https://doi.org/10.1109/TASE.2006.877401} {\bibfield  {journal} {\bibinfo
  {journal} {IEEE Transactions on Automation Science and Engineering}\ }\textbf
  {\bibinfo {volume} {3}},\ \bibinfo {pages} {360} (\bibinfo {year}
  {2006})}\BibitemShut {NoStop}%
\bibitem [{\citenamefont {Yadav}\ and\ \citenamefont
  {Sharma}(2023)}]{REV_ML_SNL}%
  \BibitemOpen
  \bibfield  {author} {\bibinfo {author} {\bibfnamefont {P.}~\bibnamefont
  {Yadav}}\ and\ \bibinfo {author} {\bibfnamefont {S.~C.}\ \bibnamefont
  {Sharma}},\ }\bibfield  {title} {\bibinfo {title} {A systematic review of
  localization in wsn: Machine learning and optimization-based approaches},\
  }\href {https://doi.org/https://doi.org/10.1002/dac.5397} {\bibfield
  {journal} {\bibinfo  {journal} {International Journal of Communication
  Systems}\ }\textbf {\bibinfo {volume} {36}},\ \bibinfo {pages} {e5397}
  (\bibinfo {year} {2023})}\BibitemShut {NoStop}%
\bibitem [{\citenamefont {Pierangeli}\ \emph {et~al.}(2021)\citenamefont
  {Pierangeli}, \citenamefont {Rafayelyan}, \citenamefont {Conti},\ and\
  \citenamefont {Gigan}}]{Claudio_2021}%
  \BibitemOpen
  \bibfield  {author} {\bibinfo {author} {\bibfnamefont {D.}~\bibnamefont
  {Pierangeli}}, \bibinfo {author} {\bibfnamefont {M.}~\bibnamefont
  {Rafayelyan}}, \bibinfo {author} {\bibfnamefont {C.}~\bibnamefont {Conti}},\
  and\ \bibinfo {author} {\bibfnamefont {S.}~\bibnamefont {Gigan}},\ }\bibfield
   {title} {\bibinfo {title} {Scalable spin-glass optical simulator},\ }\href
  {https://doi.org/10.1103/PhysRevApplied.15.034087} {\bibfield  {journal}
  {\bibinfo  {journal} {Phys. Rev. Appl.}\ }\textbf {\bibinfo {volume} {15}},\
  \bibinfo {pages} {034087} (\bibinfo {year} {2021})}\BibitemShut {NoStop}%
\bibitem [{\citenamefont {Lagoudakis}\ and\ \citenamefont
  {Berloff}(2017)}]{Lagoudakis_2017}%
  \BibitemOpen
  \bibfield  {author} {\bibinfo {author} {\bibfnamefont {P.~G.}\ \bibnamefont
  {Lagoudakis}}\ and\ \bibinfo {author} {\bibfnamefont {N.~G.}\ \bibnamefont
  {Berloff}},\ }\bibfield  {title} {\bibinfo {title} {A polariton graph
  simulator},\ }\href {https://doi.org/10.1088/1367-2630/aa924b} {\bibfield
  {journal} {\bibinfo  {journal} {New Journal of Physics}\ }\textbf {\bibinfo
  {volume} {19}},\ \bibinfo {pages} {125008} (\bibinfo {year}
  {2017})}\BibitemShut {NoStop}%
\bibitem [{\citenamefont {Kalinin}\ and\ \citenamefont
  {Berloff}(2018{\natexlab{a}})}]{Kalinin_2018}%
  \BibitemOpen
  \bibfield  {author} {\bibinfo {author} {\bibfnamefont {K.~P.}\ \bibnamefont
  {Kalinin}}\ and\ \bibinfo {author} {\bibfnamefont {N.~G.}\ \bibnamefont
  {Berloff}},\ }\bibfield  {title} {\bibinfo {title} {Networks of
  non-equilibrium condensates for global optimization},\ }\href
  {https://doi.org/10.1088/1367-2630/aae8ae} {\bibfield  {journal} {\bibinfo
  {journal} {New Journal of Physics}\ }\textbf {\bibinfo {volume} {20}},\
  \bibinfo {pages} {113023} (\bibinfo {year} {2018}{\natexlab{a}})}\BibitemShut
  {NoStop}%
\bibitem [{\citenamefont {Kalinin}\ and\ \citenamefont
  {Berloff}(2018{\natexlab{b}})}]{Kalinin2018_SciRep}%
  \BibitemOpen
  \bibfield  {author} {\bibinfo {author} {\bibfnamefont {K.~P.}\ \bibnamefont
  {Kalinin}}\ and\ \bibinfo {author} {\bibfnamefont {N.~G.}\ \bibnamefont
  {Berloff}},\ }\bibfield  {title} {\bibinfo {title} {Global optimization of
  spin {Hamiltonians} with gain-dissipative systems},\ }\href
  {https://doi.org/10.1038/s41598-018-35416-1} {\bibfield  {journal} {\bibinfo
  {journal} {Scientific Reports}\ }\textbf {\bibinfo {volume} {8}},\ \bibinfo
  {pages} {17791} (\bibinfo {year} {2018}{\natexlab{b}})}\BibitemShut {NoStop}%
\bibitem [{\citenamefont {Gao}\ \emph {et~al.}(2017)\citenamefont {Gao},
  \citenamefont {Latorre},\ and\ \citenamefont {Ruan}}]{gao2017book}%
  \BibitemOpen
  \bibfield  {author} {\bibinfo {author} {\bibfnamefont {D.~Y.}\ \bibnamefont
  {Gao}}, \bibinfo {author} {\bibfnamefont {V.}~\bibnamefont {Latorre}},\ and\
  \bibinfo {author} {\bibfnamefont {N.}~\bibnamefont {Ruan}},\ }\href
  {https://doi.org/10.1007/978-3-319-58017-3} {\emph {\bibinfo {title}
  {Canonical Duality Theory: Unified Methodology for Multidisciplinary
  Study}}}\ (\bibinfo  {publisher} {Springer Cham},\ \bibinfo {year}
  {2017})\BibitemShut {NoStop}%
\bibitem [{\citenamefont {Xu}\ \emph {et~al.}(2024)\citenamefont {Xu},
  \citenamefont {Chen}, \citenamefont {Hong}, \citenamefont {Fidan},
  \citenamefont {Parisini},\ and\ \citenamefont {Johansson}}]{xu_arxiv2024}%
  \BibitemOpen
  \bibfield  {author} {\bibinfo {author} {\bibfnamefont {G.}~\bibnamefont
  {Xu}}, \bibinfo {author} {\bibfnamefont {G.}~\bibnamefont {Chen}}, \bibinfo
  {author} {\bibfnamefont {Y.}~\bibnamefont {Hong}}, \bibinfo {author}
  {\bibfnamefont {B.}~\bibnamefont {Fidan}}, \bibinfo {author} {\bibfnamefont
  {T.}~\bibnamefont {Parisini}},\ and\ \bibinfo {author} {\bibfnamefont
  {K.~H.}\ \bibnamefont {Johansson}},\ }\href@noop {} {\bibinfo {title} {Global
  solution to sensor network localization: A non-convex potential game approach
  and its distributed implementation}} (\bibinfo {year} {2024}),\ \Eprint
  {https://arxiv.org/abs/2401.02471} {arXiv:2401.02471} \BibitemShut {NoStop}%
\bibitem [{\citenamefont {Cormen}\ \emph {et~al.}(2009)\citenamefont {Cormen},
  \citenamefont {Leiserson}, \citenamefont {Rivest},\ and\ \citenamefont
  {Stein}}]{clrs2009introduction}%
  \BibitemOpen
  \bibfield  {author} {\bibinfo {author} {\bibfnamefont {T.~H.}\ \bibnamefont
  {Cormen}}, \bibinfo {author} {\bibfnamefont {C.~E.}\ \bibnamefont
  {Leiserson}}, \bibinfo {author} {\bibfnamefont {R.~L.}\ \bibnamefont
  {Rivest}},\ and\ \bibinfo {author} {\bibfnamefont {C.}~\bibnamefont
  {Stein}},\ }\href
  {https://mitpress.mit.edu/9780262533058/introduction-to-algorithms/} {\emph
  {\bibinfo {title} {Introduction to Algorithms}}},\ \bibinfo {edition} {3rd}\
  ed.\ (\bibinfo  {publisher} {The MIT Press},\ \bibinfo {address} {Cambridge,
  MA},\ \bibinfo {year} {2009})\BibitemShut {NoStop}%
\end{thebibliography}%

\end{document}